\documentclass[12pt]{article}

\pdfoutput=1

\usepackage{color}
\definecolor{gris}{gray}{0.50}

\usepackage{a4wide}
\usepackage{graphics}
\usepackage{graphicx}
\usepackage{epsfig}
\usepackage{amssymb}

\newcommand{\la}{\left <}
\newcommand{\ra}{\right >}
\newcommand{\graybullet}{\color{gris} \bullet}

\begin{document}

%\title{Barchan dune corridors: field characterization and investigation of control parameters}
%\author{H. Elbelrhiti, B. Andreotti and P. Claudin}

%\affil{Laboratoire de Physique et M\'ecanique des Milieux H\'et\'erog\`enes (UMR CNRS 7636), ESPCI, Paris, France.}

%%%%%%%%%%%%%%%%%%%%%%%%%%%%%%%%%%%%%%%%%%%%

\begin{center}

{\Large \bfseries Barchan dune corridors: field characterization and investigation of control parameters}

\vspace*{0.2cm}

{\large Hicham Elbelrhiti, Bruno Andreotti and Philippe Claudin}

\vspace*{0.2cm}

Laboratoire de Physique et M\'ecanique des Milieux H\'et\'erog\`enes (UMR CNRS 7636), ESPCI, 10 rue Vauquelin 75231 Paris Cedex 05, France.

\end{center}

%%%%%%%%%%%%%%%%%%%%%%%%%%%%%%%%%%%%%%%%%%%%%%

\begin{abstract}
The structure of the barchan field located between Tarfaya and Laayoune (Atlantic Sahara, Morocco) is quantitatively investigated and compared to that in La Pampa de la Joya (Arequipa, Peru). On the basis of field measurements, we show how the volume, the velocity and the output sand flux of a dune can be computed from the value of its body and horn widths. The dune size distribution is obtained from the analysis of aerial photographs. It shows that these fields are in a statistically homogeneous state along the wind direction and present a `corridor' structure in the transverse direction, in which the dunes have a rather well selected size. Investigating the possible external parameters controlling these corridors, we demonstrate that none among topography, granulometry, wind and sand flux is relevant. We finally discuss the dynamical processes at work in these fields (collisions and wind fluctuations), and investigate the way they could regulate the size of the dunes. Furthermore we show that the overall sand flux transported by a dune field is smaller than the maximum transport that could be reached in the absence of dunes, i.e. in saltation over the solid ground.
\end{abstract}

%\begin{article}

%___________________________________________________________________
\section{Introduction}
\label{intro}

Barchan dunes form under unidirectional winds --~trade winds, mostly~-- when there is a localized source of sand. By contrast, transverse dunes form under the same wind conditions when sand, previously deposited by rivers, lakes or seas, covers the solid ground [\emph{Bagnold}, 1941; \emph{Cooke et al.}, 1993]. The initial instability mechanism forming these dunes from a flat sand bed results from a coupling between the shape of the dune, the air flow around it, the sand transport and the subsequent erosion/deposition processes. The scale of the most unstable wavelength $\lambda_m$ is related to the sand flux saturation length $\ell_s$ as $\lambda_m \sim 12 \ell_s$ [\emph{Andreotti et al.}, 2002b; \emph{Elbelrhiti et al.}, 2005; \emph{Claudin and Andreotti}, 2006]. This length is the distance the sand flux needs to relax to its equilibrium (i.e. saturated) value and is, in first approximation, proportional to the turbulent drag length as $\ell_s \sim 4.4 \frac{\rho_s}{\rho_f} d$, where $\rho_{s}$ and $\rho_{f}$ stand respectively for the densities of the grains and the fluid (air), and $d$ for the grain diameter [\emph{Andreotti}, 2004; \emph{Elbelrhiti et al.}, 2005]. The formation of a single mature barchan after this initial stage, in particular the distribution of air flow and sediment flux around the dune as well as the development of its characteristic crescentic shape, has now reached a significant level of understanding [\emph{Lancaster et al.}, 1996; \emph{Frank and Kocurek}, 1996a; \emph{Frank and Kocurek}, 1996b; \emph{Wiggs et al.}, 1996; \emph{Wiggs}, 2001; \emph{Sauermann et al.}, 2003; \emph{Kroy et al.}, 2002; \emph{Hersen et al.}, 2004; \emph{Hersen}, 2004].

Barchan dunes may be observed in a variety of places on Earth [\emph{Fryberger and Dean}, 1979; \emph{Cooke et al.}, 1993] and on Mars. They can be rather isolated but they can also be found in large fields which show rather coherent and homogeneous corridor-like tens of kilometers long structures. Within these corridors, dune size and spacing seems more or less well defined but this size is not the elementary size at which dunes form as many other non-linear mechanisms come into play. Previous field works have mainly focused on the morphology and kinematics of individual dunes [\emph{Finkel}, 1959; \emph{Long and Sharp}, 1964; \emph{Norris}, 1966; \emph{Lettau and Lettau}, 1969; \emph{Hastenrath}, 1967-1987; \emph{Slattery}, 1990; \emph{Hesp and Hastings}, 1998; \emph{Parker Gay}, 1999; \emph{Sauermann et al.}, 2000; \emph{Andreotti et al.}, 2002a]. Barchan fields have also been investigated [\emph{Lettau and Lettau}, 1969; \emph{Corbett}, 1999; \emph{Kocurek and Ewing}, 2005; \emph{Ewing et al.}, 2006; \emph{Ould Ahmedou et al.}, 2006], the most detailed and complete studies concerning La Pampa de La Joya, in the Arequipa region (Peru) [\emph{Finkel}, 1959; \emph{Lettau and Lettau}, 1969; \emph{Hastenrath}, 1967-1987]. This dune field is compared here to the longest barchan field on Earth, in the Atlantic Sahara region (Morocco). At that place, due to the shouldered profile of the shoreline next to the town Tarfaya, the beaches provide some sand that can be transported inland by the winds. The barchan field starts next to the sea and remains clearly visible for hundreds of km downwind, sometimes altered or disturbed by relief such as large depressions of several km large (sebkhas) or by other dunes coming from other sand sources.

From the phenomenological point of view, several attempts have been made to interpret dune fields in the framework of complex systems and non-linear physics. Leaving apart the details of sand transport and hydrodynamics, dunes may be seen as patterns emerging by self-organization [\emph{Werner}, 2003; \emph{Kocurek and Ewing}, 2005]. The interest of this approach also constitutes its main limitation as it focuses on the morphology of patterns. It successfully relates  the symmetries of dunes to that of the wind regime [\emph{Werner}, 1995] but fails to address questions regarding for instance their time and length scales, for which it is essential to understand the dynamical mechanisms at the scale of the dune.

Barchan field stripes have been simulated by \emph{Lima et al.} [2002] assuming that the sand flux balance at the scale of a single dune makes it stable. However, it has been shown later that this stability assumption is not self-consistent [\emph{Hersen et al.}, 2004], and that, in response to perturbations such as strong wind variations or collisions, barchans destabilize toward the generation of waves on their back and flanks [\emph{Elbelrhiti et al.}, 2005], possibly leading to some calving processes.

Our aim is here two-fold. First, we wish to evidence the selection of the dune size and the corridor structure. This has been previously inferred from aerial photographs in a qualitative manner, but we provide in this paper a systematic and quantitative characterisation of barchan fields. This will serve in the future as a bench-mark to test numerical models. Second, we wish to investigate the parameter controlling the size of the dunes. Is there a difference of grain size between a zone of large dunes and a zone of small ones  [\emph{Lancaster}, 1982]? Is there a difference of wind speed? Is there a change of the nature of the soil (rock, vegetation, etc) or of the topography? Is the dune size controlled by the overall sand flux at the origin of the barchan field (the flux of sand deposited on the beach by the ocean in the case of Atlantic Sahara)? Does the dune size result from that at which they initially appeared, which would mean that the entrance conditions are memorised all along the dune field? We show that all these possibilities should be rejected, in favour of a regulation of the dune size by dynamical mechanisms. We finally discuss the hypothesis of a control of dune size by the lateral sand transport stirred during storms.
\begin{figure*}[t!]
\includegraphics{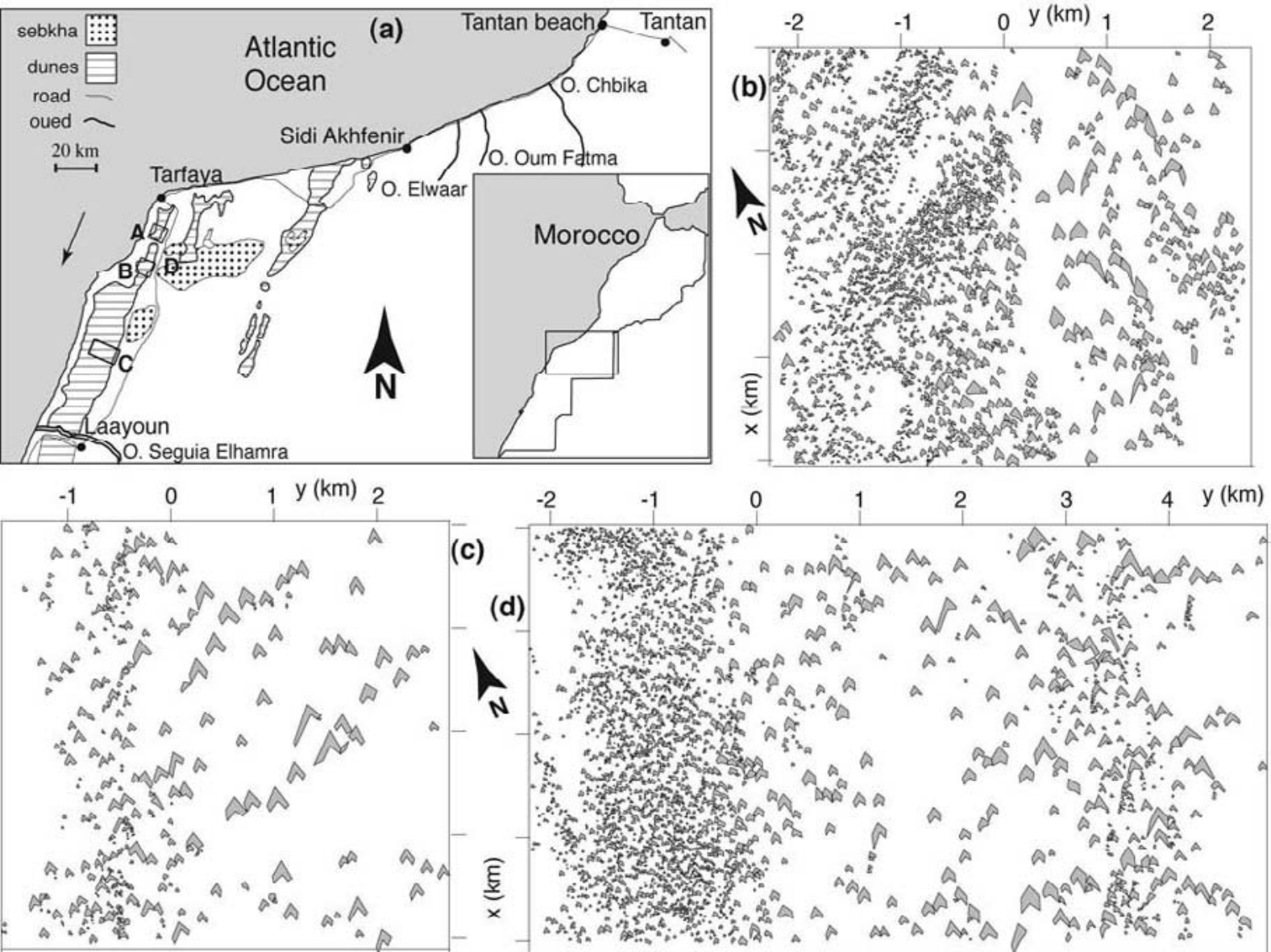}
\caption{(a) Map of Atlantic Sahara (inset) with location of the different zones of study. (b), (c) and (d): Schematic view of the barchan fields in the three zones \textsf{A}, \textsf{B} and \textsf{C} respectively. Zone \textsf{D} is the place of the `mono-corridor' discussed in the last section of the paper and displayed in figure~\ref{Monocouloir}. The $x$ axis is along the wind direction (arrow on the map), and $y$ is transverse. As defined more quantitatively later on, we set $y=0$ at the transition line between small and large dunes}
\label{3zones}
\end{figure*}
%

%___________________________________________________________________
\section{Barchan dune corridors}
\label{corridors}

In this section, we shall first give an overall and qualitative description of the two studied sites. We then show that an effective dune size selection is at work in these fields, resulting into a single or multi-corridor structure. This description is followed by the analysis of the size distribution and the quantitative evidence for the structure transverse to the wind direction.

%__________
\subsection{Two sites of study}
Two barchan fields have been studied for this work. The first is located in Atlantic Sahara (south of Morocco) where, besides collecting aerial photographs, we carried out many field investigations. The second one is a barchan field in the Arequipa region of Peru. We have not actually been there but worked with the available published documents as well as aerial views that can be found online.

\begin{figure*}[t!]
\includegraphics{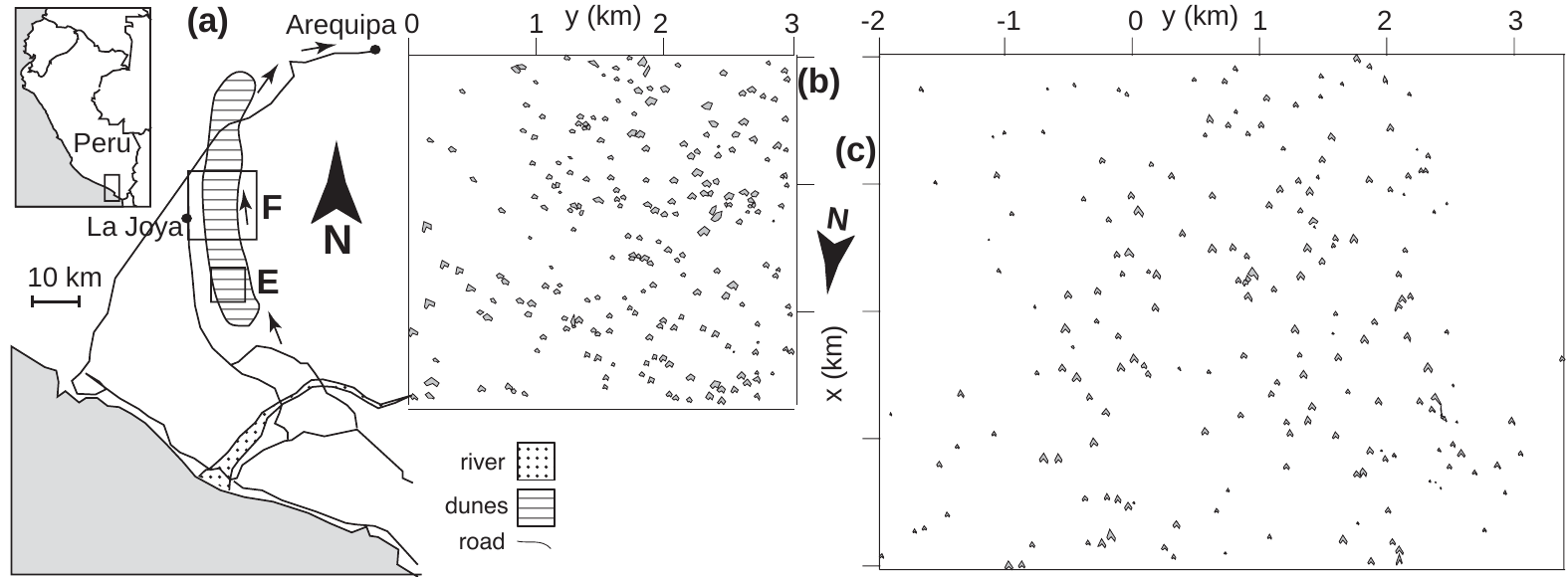}
\caption{(a) Map of Peru (inset) with location of La Joya and the two zones of study. (b) and (c): Schematic view of the barchan fields in these two zones \textsf{E} and \textsf{F} respectively. Note the change of wind direction (arrows) from La Joya to Arequipa.}
\label{JoyaMap}
\end{figure*}

A map of the Moroccan region of interest is shown in figure~\ref{3zones}a. There, the topography is very flat in the form of a limestone-sandstone (Moghrebian) plateau over which trade winds blow regularly, with a typical resultant drift potential $RDP/DP \sim 0.9$ [\emph{Elbelrhiti et al.}, 2005], i.e. a very unimodal wind regime (see Appendix~\ref{AppA}). The direction of these winds is slightly oriented to the west, making an angle of $23^\circ$ to the north. This direction is precisely the direction of dune propagation and consequently that of the corridors. We chose to study dunes in the sub-region between the towns of Tarfaya and Laayoune, which are approximately a hundred km apart. More precisely, we selected three different places: zone \textsf{A} slightly south of Tarfaya ($27^\circ51'$N, $12^\circ55'$W); zone \textsf{B} next to the town of Tah ($27^\circ42'$N, $12^\circ59'$W); zone \textsf{C} slightly north of Laayoune ($27^\circ19'$N, $13^\circ11'$W). The zones are typically $4$~km long and $4$ to $7$~km wide. The distance between \textsf{A} and \textsf{B} is about $20$~km, and $40$~km between \textsf{B} and \textsf{C}. Schematic top views of these three zones are displayed in figure~\ref{3zones}.
\begin{figure}[t!]
\includegraphics{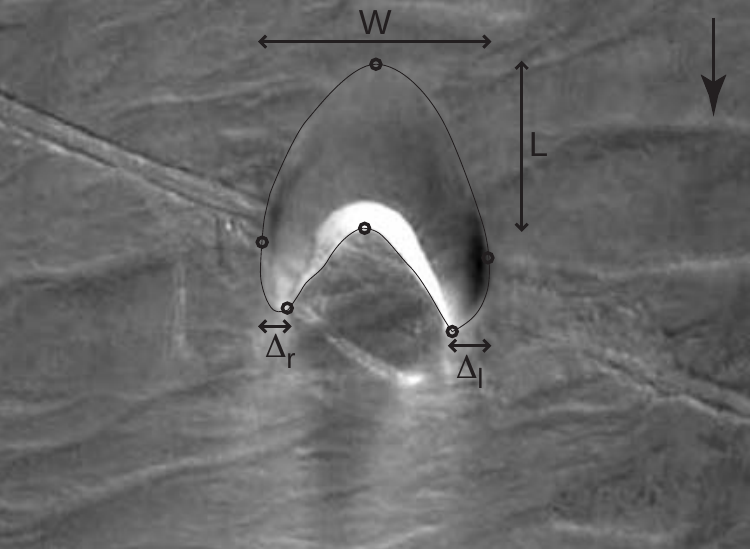}
\caption{Position of the six reference points marked to measure dune and horn widths and lengths as well as the asymmetry. This photo is from the barchan field of La Joya. The wind comes from the top (arrow). Note the sand leak at the horn tips.}
\label{6pts}
\end{figure}

Although the three fields in figure~\ref{3zones} do not look exactly the same -- e.g. dunes are clearly larger in zone \textsf{B} than in \textsf{A} and \textsf{C} -- they all show a similar structure transverse to the wind direction: dunes next to the eastern border of the field are large and far apart from each other, whereas further into the field the dunes are smaller and their spatial distribution denser. As they have a transverse expansion much smaller than along the wind direction, we call `corridor' these places of roughly uniform dune size and density. We must emphasize that the corridor pattern of zone \textsf{B} is not the direct continuation of that of zone \textsf{A} -- same with zone \textsf{C} with respect to zone \textsf{B} -- as some topographic interruptions (depressions or Sebkhas, additional sources of sand [\emph{Oulerhi}, 1992]) locally break or modify the barchan field in between these zones. However, each time the dune field runs over a large uniform terrain, the same general pattern is recovered.

It is interesting to compare these data with those from the barchan field close to La Joya railway station ($16^\circ43'$S, $71^\circ51'$W), see figure~\ref{JoyaMap}. Although this field is not very large -- approximately $30$~km long -- several studies have been devoted to this place, and barchans have been followed over decades [\emph{Finkel}, 1959;\emph{Lettau and Lettau}, 1969; \emph{Hastenrath}, 1967-1987]. The wind there blows roughly from south to north. Unfortunately, no permanent weather station has been installed in the dune field so far. The closest one is at the airport of Arequipa, close to moutains and it gives a low RDP/DP factor of the order of $0.4$, due to frequent reversing winds (see Appendix~\ref{AppA} and figure~\ref{storms}). Two zones there have been selected. A first one (\textsf{E}) close to the beginning of the field, and a second one further north close to the train station (\textsf{F}) which exactly corresponds to Lettau and Lettau's work site [\emph{Lettau and Lettau}, 1969]. These two zones are separated by $10$~km and barchans eventually disappear $20$~km further north of zone \textsf{F} in an agricultural area. A schematic view of the barchans in these zones is displayed in figure~\ref{JoyaMap}. In contrast to Atlantic Sahara, the field is very homogeneous in size and density, with a single stripe of small and diluted dunes.

%__________
\subsection{Quantitative evidence for a transverse structure}
\label{dunesizedistri}
With a careful analysis of the aerial photographs, one can get a systematic measure of some of the dune characteristics, namely the width $W$, the length $L$, and the left and right horn widths $\Delta_{l,r}$. These quantities can be computed from six data points: the back of the dune, the two sides, the two tips of the slip face at the horns, and finally the most interior point of the slip face, see figure~\ref{6pts} and Appendix~\ref{AppA}. Comparing the left and right horn widths, the asymmetry of the dune can be also estimated. For the following graphs, an approximate number of $5000$ dunes have been measured in the Moroccan field and several hundreds in Peru.

At this stage, we shall briefly discuss an important length scale issue. As a matter of fact, most of the results presented here are plotted in real units -- lengths in m, volumes in m$^3$ -- but it is essential to be able to fairly compare some data with others obtained in different situation, e.g. in another dune field. For example, what do we mean by a `small' or a `large' barchan? We have shown in previous papers that the only length-scale of the problem is the saturation length $\ell_s$, which controls for instance the size $\lambda_m\sim 12 \ell_s$ at which dunes form but also the cut-off wavelength $\lambda_c \sim 6.3 \ell_s$ below which dunes disappear [\emph{Andreotti et al.}, 2002b; \emph{Elbelrhiti et al.}, 2005; \emph{Claudin and Andreotti}, 2006]. For the grain size $d \sim 175~\mu$m found in Morocco (see figure~\ref{ControlParams}), the saturation length has been directly measured ($\ell_s \sim 1.7$~m), which gives $\lambda_m \sim 20$~m and $\lambda_c \sim 11$~m. This is consistent with the wavelength of the surface undulations which decorate the large barchans as well as the length of the smallest barchans observed in the region [\emph{Elbelrhiti et al.}, 2005]. On Mars however, $\lambda_m$ is of the order of $600$~m and so is the typical size of the smallest dunes there [\emph{Claudin and Andreotti}, 2006]. Therefore, although for the sake of readability we did not make all our plots dimensionless, lengths should be implicitly thought in units of $\lambda_m$ (or $\ell_s$). Finally note that the sand grains of the barchan field of La Joya have $d \sim 150~\mu$m and $\rho_s \sim 2300$~kg/m$^3$ [\emph{Hastenrath}, 1967-1987] which give $\lambda_m \sim 15$~m, a slightly smaller value than in the field of the Atlantic Sahara.

\begin{figure*}[t!]
\includegraphics{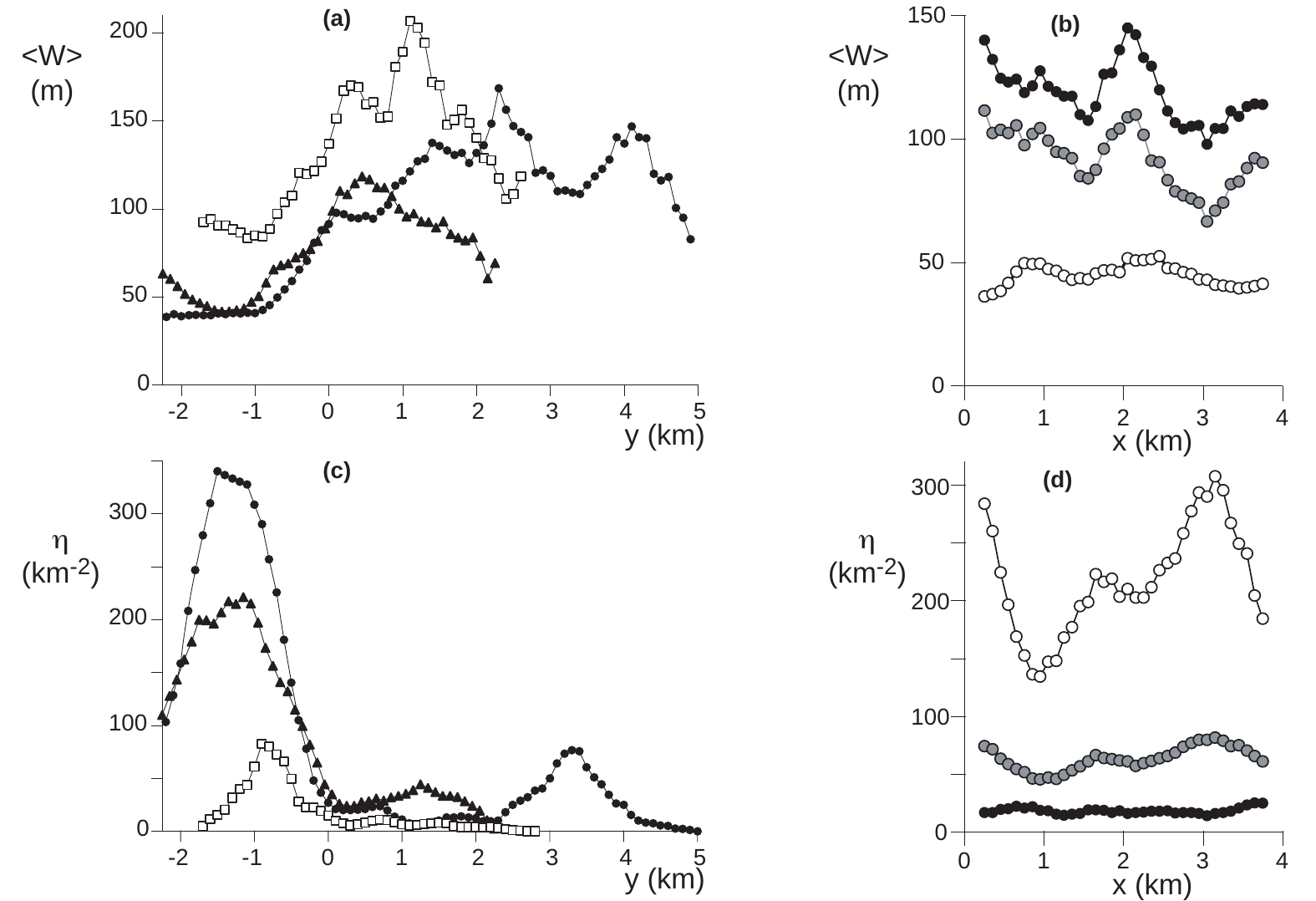}
\caption{(a) Mean dune width $\la W \ra$ and (c) number $\eta$ of dunes per km$^2$ as a function of the transverse position $y$ in the three Moroccan zones. For comparison, the corresponding data computed from the barchan field of La Joya are displayed in figure~\ref{Joyadata}. In all these graphs, {\large $\blacktriangle$} is for zone \textsf{A}, {\small $\square$} is for zone \textsf{B}, and {\Large $\circ$}, {\Large $\graybullet$}, {\Large $\bullet$} for zone \textsf{C}. As evidenced by the packing density profile, $y=0$ is the transition point between the small ($y<0$) and large ($y>0$) dune corridors. Panels (b) and (d) show respectively profiles of the mean width and the dune density along the wind direction ($x$) for zone \textsf{C}. White symbols represent an analysis restricted to the small-dune corridor ($y<0$), black symbols are for the large-dune one ($y>0$) and there is no restriction on $y$ for the grey symbols.}
\label{versusy}
\end{figure*}

In order to reveal the transverse structure of the dune field, we investigate rectangular working domains $\mathcal{A}$ of area $\delta_x \times \delta_y$, whose long side is along the wind ($x$ direction) and of the size of the whole zone of study ($3$ to $4$~km, see figures~\ref{3zones} and ~\ref{JoyaMap}), and whose short side is a small fraction of the transverse dimension, typically $\sim 500$~m, centered at some value of coordinate $y$.  In figure~\ref{versusy}a,b is displayed the mean dune width,
\begin{equation}
\la W \ra = \int \! W \, P_2 (W,\mathcal{A}) \, dW =
\frac{1}{N_\mathcal{A}} \times
\frac{\sum_{i \in \mathcal{A}} W_i^3}{\sum_{i \in \mathcal{A}} W_i^2} \, ,
\label{def_meanW}
\end{equation}
defined from the dune size probability density function (PDF) weighted by the surface of the dune, $P_2 (W)$. The other possible weighting options for the definition of such PDFs are discussed in Appendix~\ref{AppB}. Figure~\ref{versusy}c,d shows the dune density (number of dunes per unit surface -- here in km$^2$),
\begin{equation}
\eta= \frac{N_\mathcal{A}}{\delta_x \delta_y} \, ,
\label{def_eta}
\end{equation}
as a function of $y$. In all three Moroccan zones, one can clearly see that in the western part of the field the dunes are smaller and more densely packed. The location of the passage from one corridor to the other can be determined by looking at the shape of the curve $\eta(y)$. We can indeed set $y=0$ at the toe of the large left peak. $y<0$ is then the domain of small dunes, and conversely $y>0$ is that of large ones.

\begin{figure*}[t!]
\includegraphics{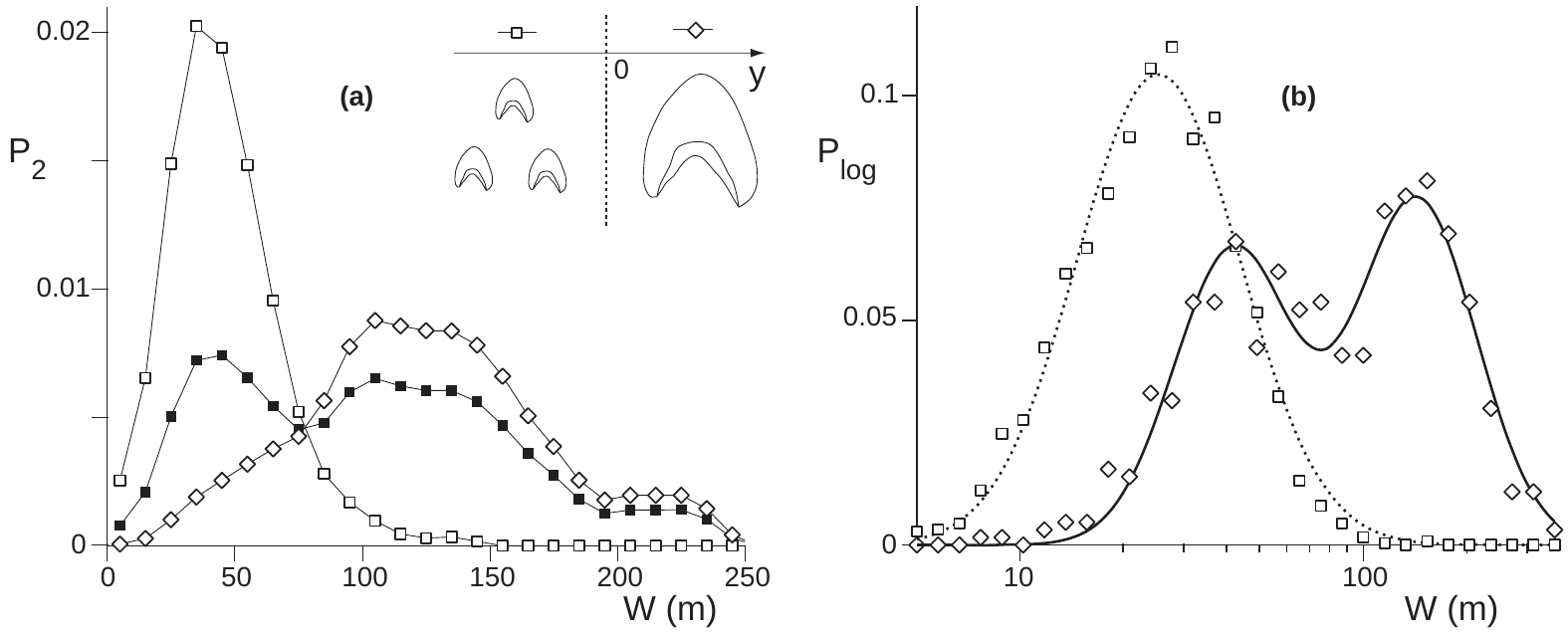}
\caption{(a) Dune width probability density functions $P_2(W)$ computed over the Laayoune field --~see schematic inset or the general view of zone \textsf{C} in figure~\ref{3zones}d. The PDFs, computed with a bin width $\delta W \sim 10$~m, compares the statistics over the whole zone \textsf{C} ({\small $\blacksquare$}), to that over the corridors of small ({\small $\square$}, $y<0$) and large ($\lozenge$, $y>0$) barchans. (b) Probability distribution function $P_{\rm log}$ of the dune width logarithm, weighted in number of barchans, for the small ({\small $\square$}, $y<0$) and large ($\lozenge$, $y>0$) dune corridors. In the western zone $y<0$, the distribution presents a single peak, well fitted by a log normal distribution (dotted line), which corresponds to the size at which dunes appear (peak at $W \sim 25$~m). By contrast, in the eastern zone $y>0$, the distribution presents a first peak corresponding to large dunes but also a second one corresponding to the small dunes calved from the large ones. The solid line is the best fit by the sum of two log-normal distributions. The two peaks are located at $W \sim 42$~m and $W \sim 145$~m.}
\label{histos}
\end{figure*}

A more careful inspection of graphs (a) and (c) of figure~\ref{versusy} shows that, although the three zones share qualitatively the same corridor pattern, there are quantitative differences between them. The small dunes of zone \textsf{B} are for instance almost of the size of the large ones in zone \textsf{A}. Another example is that the small dunes of zones \textsf{A} and \textsf{C} are of similar size, but there is a $50\%$ increase in the density from the former to the latter. Finally, a secondary peak in the $\eta(y)$ profile of zone \textsf{C} is visible precisely at the position where the corresponding curve $\la W \ra \! (y)$ drops down, showing how correlated these two quantities are.

Furthermore, we have investigated the homogeneity in size and density along the windward direction $x$. In the same spirit as for the $y$ profiles, rectangular working domains are chosen with a short size $\delta_x=500$~m, and a long side $\delta_y$ which can cover either the whole field, or restricted to the sub-zones $y<0$ (small-dune corridor) or $y>0$ (large-dune corridor). The corresponding profiles $\la W \ra \! (x)$ and $\eta (x)$ are displayed in figure~\ref{versusy}b,d. The data corresponding to the small-dune corridor appear to be very homogeneous (flat profiles) in dune size, but quite fluctuating in density. Looking at the field in figure~\ref{3zones}d, one can indeed identify a hole of dunes on the western side, followed by a renucleation of barchans further downwind. Large dunes are more dispersed in size, although without any particular trend as a function of $x$. However, their density is constant all along the corridor.
\begin{figure}[t!]
\includegraphics{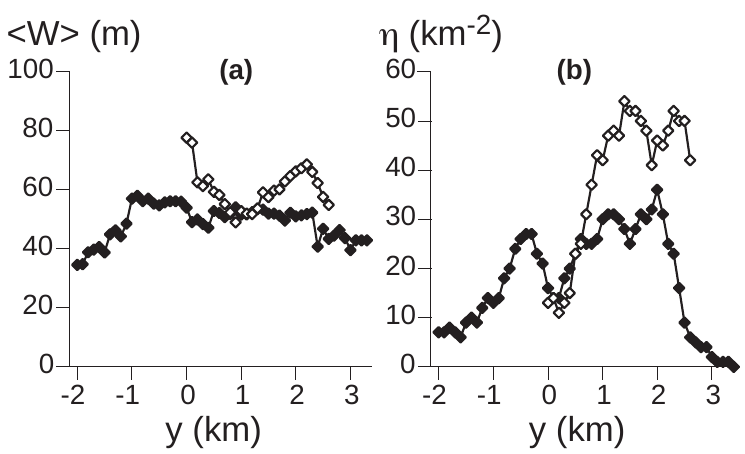}
\caption{Transverse profiles of the mean dune width (a) and dune density (b) in the two zones of La Joya barchan field. $\lozenge$ is for zone \textsf{E} and $\blacklozenge$ for zone \textsf{F}.}
\label{Joyadata}
\end{figure}

A richer quantity than the average dune width is the entire dune width distribution. Figure~\ref{histos}a shows this distribution for the whole zone \textsf{C}, as well as for its small and large dune corridors. The size distributions are respectively peaked at $W \sim 40$~m ($\sim 2 \lambda_m$) and $W \sim 125$~m ($\sim 6 \lambda_m$) in the western and eastern corridors, with a typical standard deviation to mean ratio of the order of unity. This means that the dominant type of dunes is clearly distinct in the eastern and western corridors, in terms of surface occupied or sand volume. Figure~\ref{histos}b, which shows the PDF of the dune width logarithm, allows to refine this analysis. As it is weighted in number of dunes (see Appendix~\ref{AppB}), it emphasizes the presence of small dunes in the large dunes corridor. In the western zone, the distribution presents a single narrow peak at $W \sim 25$~m ($\sim \lambda_m$). Conversely, in the eastern zone, the distribution is wider and presents two distinct peaks corresponding to the dominant species (around $W \sim 145$~m, or $\sim 7 \lambda_m$) and to small dunes calved from the large ones (around $W \sim 42$~m, or $\sim 2 \lambda_m$). This clearly demonstrates that the size selection does not result from scale free processes.

In the barchan field of la Pampa de La Joya, the dune width distribution presents a single peak around $W \sim 50$~m ($\sim 3 \lambda_m$), also with a typical standard deviation of the same order of magnitude. Dune width and density transverse profiles are plotted in figure~\ref{Joyadata}a,b. The comparison of zones \textsf{E} and \textsf{F} shows that the field becomes wider, less dense and more homogeneous downwind. The dune density is ten times smaller than in the area of the Moroccan field where dunes have a similar size. The  `equivalent sand height' introduced by \emph{Lettau and Lettau} [1969] provides a particularly striking illustration of this difference. If one could spread all the volume $V_i$ of sand contained in the dunes, it would form a uniform layer of height $H_e = \frac{1}{\delta_x \delta_y} \sum_{i \in \mathcal{A}} V_i$. In La Joya field, one finds $H_e \sim 2$~cm [\emph{Lettau and Lettau}, 1969] whereas $H_e \sim 50$~cm in the Moroccan one.

From the previous figures, it then appears that the barchan size looks well selected, suggesting that the dunes have reached a statistical equilibrium state controlled by some parameter(s) which varies along the direction transverse to the wind. Before investigating these control parameters, we need to discuss in detail the measurement and the distribution of sand fluxes inside a dune field.

%___________________________________________________________________
\section{Sand fluxes}

A difficult, but crucial, quantity to measure in this context is the total sand flux $Q_t$ transported by the dune field. Following \emph{Lettau and Lettau}, [1969], two terms of $Q_t$ must be discussed independently, namely the `bulk' $Q_b$, and the `free' $Q_f$ fluxes. The former is related to the sand transported by the dunes themselves, whereas the latter corresponds to the grains moved in saltation by the wind in between the barchans.

\begin{figure}[t!]
\includegraphics{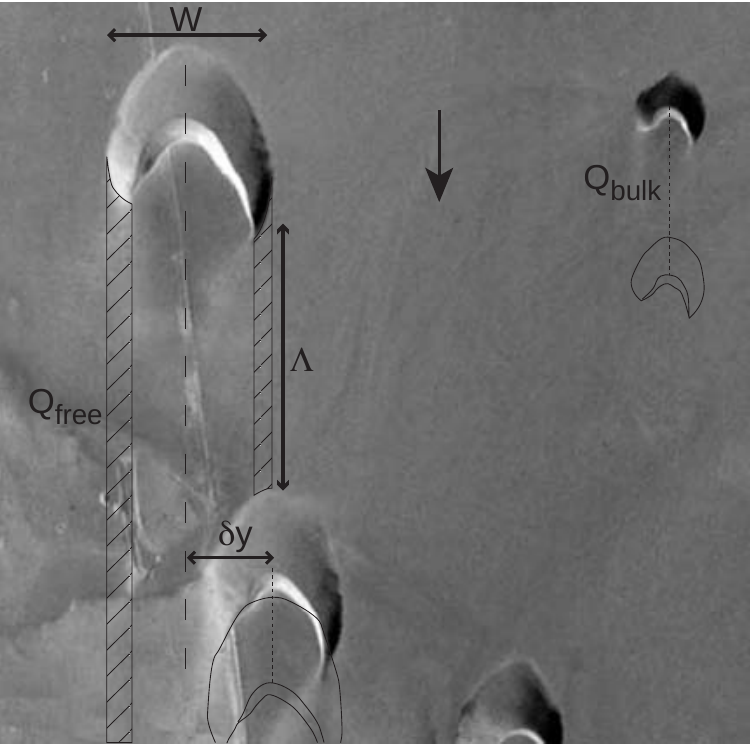}
\caption{Schematic illustration of the measurement method of both bulk and free fluxes. This photo is from the barchan field of La Joya. The wind comes from the top (arrow). $\Lambda$ is the distance from the horn tip of a barchan to the nearest downwind dune. For the purpose of figure~\ref{canard}, we note $\delta y$ the lateral distance between the centers of a dune and its next downwind neighbor.}
\label{schemafluxes}
\end{figure}

\subsection{Bulk flux}
In the domain $\mathcal{A}$ of size $\delta_x$ and $\delta_y$ in directions $x$ and $y$ respectively, we define in accordance with [\emph{Lettau and Lettau}, 1969] the volumetric bulk flux as
\begin{equation}
Q_b = \frac{1}{\delta_x \delta_y} \sum_{i \in \mathcal{A}} c_i V_i,
\end{equation}
where $c_i$ is the velocity of the dune $i$, and $V_i$ its volume. We show in the appendix~\ref{AppA} that both volume and velocity can be computed as functions of the dune width $W$, so that $Q_b$ can thus be computed directly from the analysis of aerial photographs. It should be emphasized that the dune propagation speed is proportional to the saturated flux over a flat bed $Q$, so that all wind dependance is encoded into this factor

%__________
\subsection{Free flux}
The evaluation of the free flux $Q_f$ is much less straight forward and requires more assumptions. In Lettau and Lettau's paper, $Q_f$ (called `streamer' flux in [\emph{Lettau and Lettau}, 1969]) is computed in the following manner. They define the coefficient $r$ to be such that $Q_b = r Q_t$. It represents the fraction of intercepted streamers and has been estimated between $0.4$ and $0.6$ by these authors, taking the ratio of the total width $\sum_{i \in \mathcal{A}} W_i$ (integrated over a transverse strip $\mathcal{A}$ of windward extent $600$~m) to the width of the overall lateral width of the barchan field. A better estimation may be obtained with $r \sim \la W \ra^2 \! / \eta$. The values $\la W \ra \sim 50$~m and $\eta \sim 30$ dunes per km$^2$ that can be read in figure~\ref{Joyadata} lead to $r$ of the order of $0.1$. With $Q_t = Q_b + Q_f$, it finally gives $Q_f=(1-r)Q_t=\frac{1-r}{r} \, Q_b$.

This proportionality relationship between the fluxes is a strong assumption which is far from being obvious. We would like to relax it and find a way to measure $Q_f$ independently of $Q_b$. We remark that, as the recirculation bubbles in the lee side of barchans act as perfect sand traps, all grains in saltation between the dunes must have originally escaped a barchan from one of its horns. From numerical simulations [\emph{Hersen et al.}, 2004], it can be inferred that the flux on the horns is proportional to the saturated sand flux on a flat bed $Q$, with a pre-factor close to unity. This is consistent with the measurement of the wind velocity in the horns which is found to be close to that far from the dune. In other words, the flux is saturated and the horns are flat. Assuming that all the escaping flux reaches the next downwind dune, $Q_f$ can be expressed by the means of matter conservation as:
\begin{equation}
Q_f = \frac{Q}{\delta_x \delta_y} \sum_{i \in \mathcal{A}}
\left ( \Delta_i^l \Lambda_i^l + \Delta_i^r \Lambda_i^r \right ),
\label{qfree}
\end{equation}
where $\Lambda_i^{l,r}$ is the distance from the dune $i$ to its nearest neighbor downwind starting from the tip of its left/right horn, and $\Delta_i^{l,r}$ the width of that horn. The method of measurement of both $Q_b$ and $Q_f$ is illustrated in figure~\ref{schemafluxes}.

It is important to discuss at this stage the several sources of errors or uncertainties in this measure of $Q_f$. The first one is the precision at which we were able to locate the dune reference points on the aerial photographs (see figure~\ref{6pts}). As a matter of fact, some of the photos we analyzed are very sharp with nice lightings and shadows accuentuating the edge points of the slip face. Other views are less contrasted and these positions are less well defined. As a result, the typical error in the direct determination of the $\Delta$'s is of the order of $10\%$. A second source of uncertainty is the fact that the value of the sand flux at the barchan horns has not been measured directly, but comes from the result of a model as well as qualitative observations. This would probably count for another $10$ to $20\%$. A third point to emphasize is that, as aerial photos of barchan field only give a snapshot of the state of the dunes, such an analysis gives an `instantaneous' value for $Q_f$, corresponding to the date at which the view was taken. This remark makes particular sense when comparing the zones \textsf{E} and \textsf{F} of the la Joya field (see figure~\ref{DeltaofW}) which we shall come back to later. The computation of $Q_b$ therefore contrasts with that of the bulk flux which is based on time and ensemble averaged laws for the dune volume and velocity as a function of the dune width (see apendix). A final source of error, which is difficult to evaluate, is related to the possibility of the erosion of the lime-sandstone ground on which the barchans live. In fact, a local erosion rate of the order of a fraction of a mm per year would not affect our estimation of $Q_f$ as the correspoding additional sand flux reaching the back of a dune would be this rate multiplied by the typical distance between two neighboring dunes, i.e. tens to hundreds of meters and is therefore negligible in comparison to the measured value of $Q\sim 100$~m$^2$/year (see Appendix~\ref{AppA}).

Another important point concerns the upwind boundary conditions in the $Q_f$ calculation. For each dune of a selected domain in a given photo, its next downwind neighbors are located. However, the most upwind dunes are also bombarded by sand grains, but coming from dunes out of the domain. We then run this neighbor searching routine not only on the selected domain, but all over the photograph. In the case of La Joya, the dune density is so low that some of the dunes providing sand at the upwind boundary are not actually visible on the photograph. As these dunes actually exist and are of the same size as those visualized, we have added at the upwind edge of the photograph `false' barchans as sand flux sources, duplicated from real ones. In the worse case (close to La Joya station), this procedure affects the result in the upper third of zone \textsf{F}. However, as the field is observed to be homogeneous, this should not be the dominant source of error.

\begin{figure*}[t!]
\includegraphics{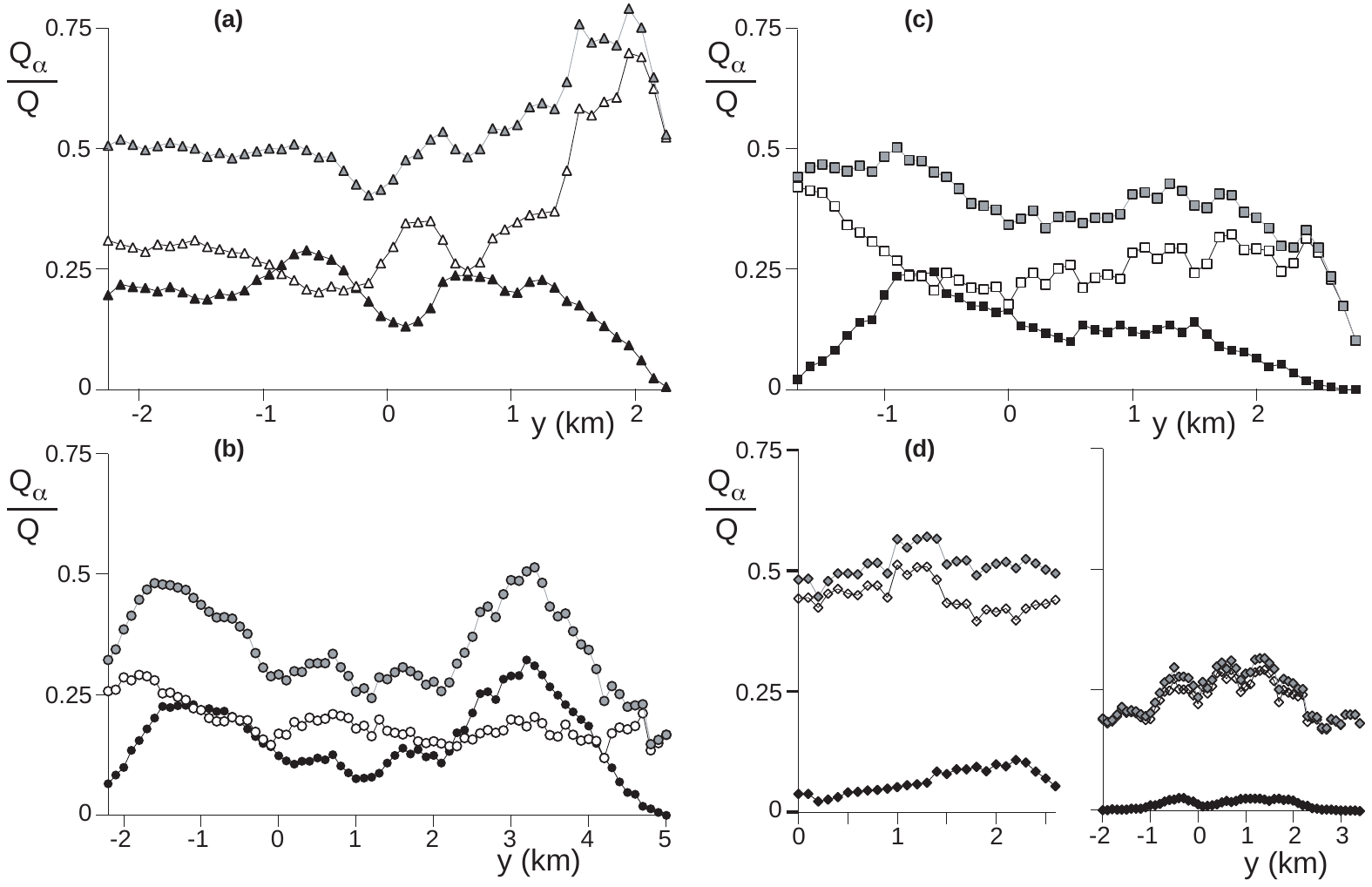}
\caption{Transverse profiles of the bulk ($Q_b$, black symbols), free ($Q_f$, white symbols) and total ($Q_t=Q_b+Q_f$, grey symbols) fluxes for the three Moroccan zones \textsf{A}, \textsf{B} and \textsf{C} in (a), (b) and (c) respectively. Same in panel (d) with La Joya barchan field: zones \textsf{E} (left) and \textsf{F} (right). All these fluxes are in units of $Q$, the saturated sand flux over a flat bed.}
\label{FreeBulkTot}
\end{figure*}
%

%__________
\subsection{Total flux}
In figure~\ref{FreeBulkTot}a,b,c are displayed the bulk, free and total fluxes as a function of the transverse coordinate $y$ for the Moroccan dunes. Interestingly, both $Q_b$ and $Q_f$ are of the same order of magnitude, but not uniform across the field. More precisely, far from being proportional to each other, a decrease of $Q_b$ in a dune `hole' is correlated with an increase of $Q_f$ -- see e.g. panel (a) around $y=0$. In total, the $Q_t$ data display rather flat profiles \emph{versus} $y$. Furthermore, the total flux value is pretty much the same in the three zones of study, indicating that $Q_t$ is quite uniform in the windward direction over long distances. Finally, this value is only a fraction (roughly one half) of the saturated flux $Q$. This means that this amount of sand could well be transported in saltation over the solid ground. This as a striking consequence. Consider a flat zone where a given sand flux is transported in saltation. If the flux is smaller than its saturated value, the system remains in this state. If, under the same conditions, some dunes acting as finite amplitude perturbations are introduced, they may survive, leading to a dune field transporting, in the bulk of dunes and in inter-dune free flux, the same overall amount of sand. We shall illustrate this transition from one state to the other in the last section.

%__________
\subsection{Sand flux balance at the scale of the field}
In the introduction, we have raised the question of the parameters controlling the size of dunes in a given area. A distinct but related issue is the global sand balance at the scale of the dune field. Does all the sand of the field come from a unique source at the upwind edge (a beach in Morocco and a less well defined mountainous place in the case of La Joya)? Is there a sand leak at the lateral boundaries? Is there significant erosion of the ground providing an additional source of sand for the dunes? Why does the Moroccan field remain collimated over such a long distance?

Within zone \textsf{F}, \emph{Lettau and Lettau} [1969] came to the conclusion that there must be a slight downwind increase of sand transport, this increase being interpreted as the signature of a local erosion of the ground ($0.2$~mm/year). In view of our results for zones \textsf{E} and \textsf{F} (figure~\ref{FreeBulkTot}d), the validity of this result can be seriously questioned. Indeed,  the dune density is so small that the total flux is dominated by the free flux, badly estimated in [\emph{Lettau and Lettau}, 1969] and quite imprecise here also.

Of course, the hypothesis of a local erosion rate as large as one grain diameter per year cannot be rejected but can neither be proved on this basis. If this was the case, integrated over a tens of kilometers, the additional flux would be larger than the entrance flux. To explain the downstream homogeneity (in particular in Morocco), the extra sand source due to erosion would need to be compensated by a lateral sand leak. We can imagine two different processes for this transverse flux. First, the grains in saltation experience random rebounds on the soil, leading to some lateral diffusion. A random-walk-like analysis predicts that, for a windward propagation $L_x$, a typical transverse displacement $L_x \sim \sqrt{D L_x}$ is expected, where $D$ is the typical lateral displacement of a salton after one hop. A reasonable estimate is $D \sim 0.1$~m, which means that $L_y$ is only about several tens of meters for $L_x=10$~km. In other words, such a diffusion-like process cannot be responsible for a broadening of the corridor strip. Second, when the wind direction changes significantly (either due to the gentle daily variation or during a storm), the sand escaping dunes can be transported outside the dune field. In that case, $L_y$ would scale linearly with $L_x$ itself rather than with its square root. The corresponding transverse flux would thus be a fraction of $Q$. Of course, during such events, there is also a sand supply on the other side. Moreover, during our field trips in Morocco, we never observed sand trapped behind bushes and stones (nebkhas) outside the dune field.

In conclusion, there is no direct proof that the sand flux escaping from the sea, in Morocco (from the mountain in La Joya) remains collimated and gives the essential contribution to the amount of sand in the field. However, an indirect indication of this hypothesis comes from the total length ($300$~km) of the Moroccan barchan field. It takes $\sim 10000$~years for a middle sized dune to propagate from the see over such a distance. Similarly, one of the mega-barchans, which moves at $2~$m/year, is now at $\sim 10$~km from the shoreline, giving a starting time of $\sim 5000$~years before present. These dates are reasonably consistent with the interglacial period corresponding to the formation of this desert, indicating that the major source of sand is indeed the ocean close Tarfaya.

%___________________________________________________________________
\section{Investigation of possible parameters controlling the corridor structure}
\label{param}

We have demonstrated that there is a regulation of the size of dunes inside corridors. As a general statement, there should exist at least one external control parameter varying from a corridor of large dunes to a corridor of small ones. There are in fact very few possible candidates, namely (i) the sand flux, (ii) the soil and the granulometry and (iii) the wind velocity across the dune field.
\begin{figure}[p]
\includegraphics{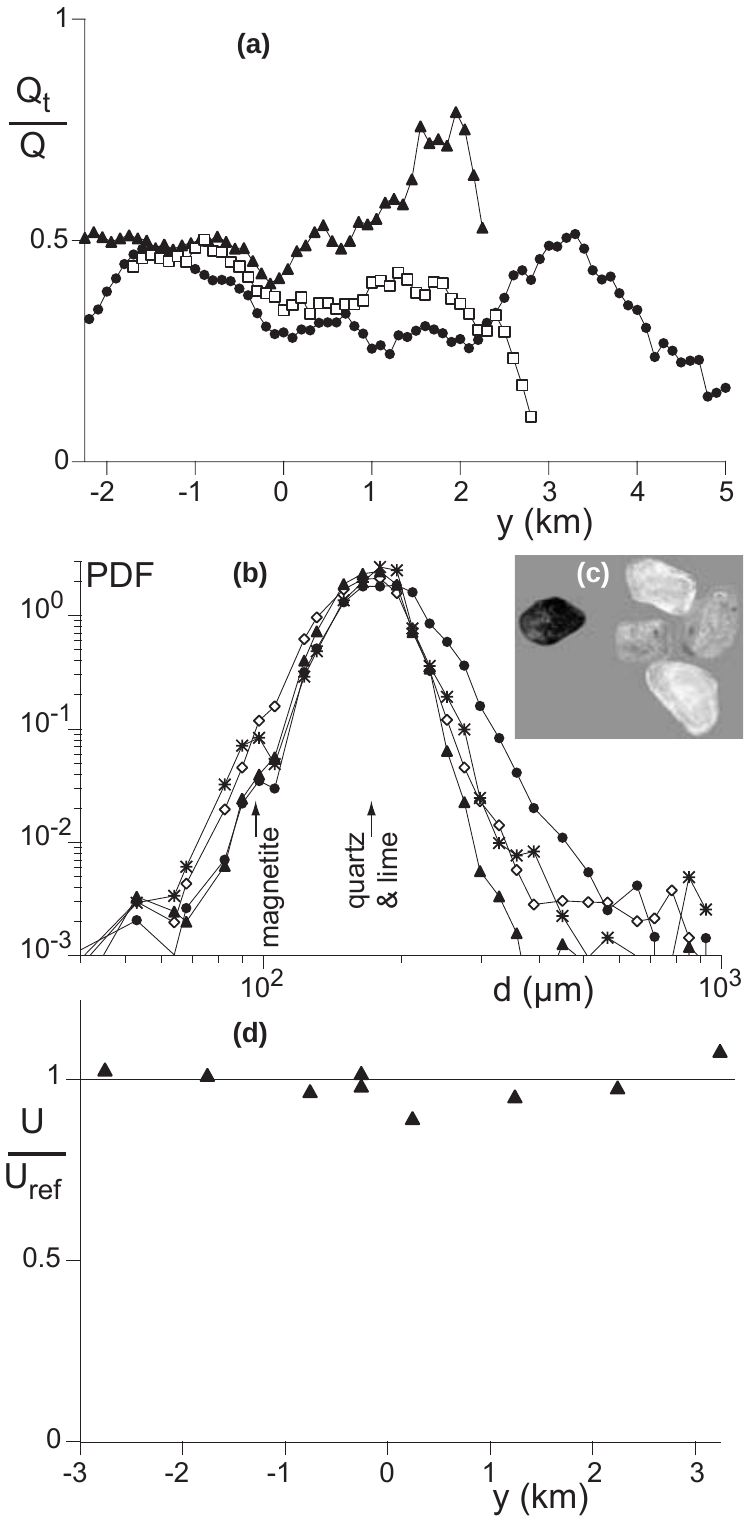}
\caption{(a) Total sand flux transverse profiles $Q_t(y)$ in zones \textsf{A} ({\large $\blacktriangle$}), \textsf{B} ({\small $\square$}) and \textsf{C} ({\Large $\bullet$}). (b) Grain size distribution. The different data sets come from sand samples taken at mid slip face of dunes, in various places: Chbika beach ($\lozenge$, $28^\circ18'$N, $11^\circ32'$W), zone \textsf{A} ({\large $\blacktriangle$}), Mega barchan ({\Large $*$}) and zone \textsf{C} ({\Large $\bullet$}). The distribution is log-normal (note the log scales on the axis) with a peak around $175~\mu$m. These grains are made of quartz and lime. A secondary peak at $100~\mu$m is clearly visible and corresponds to grains of magnetite. (c) Photograph of few grains of magnetite (black), lime from a shell (orange) and quartz (white). (d) Wind velocity across zone \textsf{A}. These wind data have been rescaled by a reference velocity measured at a fixed point. Recall $y=0$ is the transition point between the small ($y<0$) and large ($y>0$) dune corridors}
\label{ControlParams}
\end{figure}
%

%__________
\subsection{Overall sand flux}
Based on the idea that the sand flux escaping from the ocean in Tarfaya beach is conserved all along the dune field, an appealing idea is to imagine that this initial sand source is non-uniform and that, in order to transport a different quantity of sand, the system spontaneously selects a mode with an adaptated dune size. For instance, one could imagine that a field of small dunes would transport more sand than a field of large ones (or vice-versa).

The transverse profiles of the total sand fluxes $Q_t$ computed for the three Moroccan zones have been gathered in figure~\ref{ControlParams}a (see also figure~\ref{FreeBulkTot}). As already mentionned, their values are very consistent from one zone to the other and show no particular trend as a function of $y$. More precisely, one cannot indentify any clear variation of $Q_t$ around the position $y=0$ where one changes from a dune corridor to the other. We can then conclude that barchan corridors cannot be explained by transverse variations of the total sand flux.

This result can be justified \emph{a posteriori} by the following hand waving argument. If we assume that the spacing between dunes scales on the dune size ($\eta\propto1/W^2$) and that the horn size $\Delta$ is a fraction of $W$, with a velocity $c$ proportional to $1/W$ and a volume scaling as $W^3$, one gets a flux independent of the dune size. In these conditions, it would have been surprising to be able to associate a different flux to large and small dune corridors.

In the zone \textsf{E} of La Joya the same total flux  $Q_t \sim 0.5 \, Q$ as in Morocco is observed, with a very different packing density. This reinforces the idea that, although a sand flux is definitively needed to produce barchans, it neither controls their size nor their density. Finally, the comparisons between the two zones of La Joya reveals an interesting feature as the total flux drops by a factor of two from \textsf{E} to \textsf{F}. As will be discussed below, this difference should be related to the different dates at which the photographs where taken. In zone \textsf{E}, a storm has strongly affected the shape of the dunes, leading to an increase of the horn width and thus of the free flux.

%__________
\subsection{Soil and granulometry}
Another idea is to test whether the size distribution of the sand grains has a significant influence on the corridor pattern. As a matter of fact, the analysis of a wide collection of field data by \emph{Wilson} [1972], as well as the work of \emph{Lancaster} [1982] in Namibia suggest that the grain size $d$ is correlated with the size of the dunes. At a more fundamental level, we have shown how the initial wavelength of sand bed destabilization is related to the drag length $\ell_d = \frac{\rho_s}{\rho_f} d$ [\emph{Andreotti et al.}, 2002b; \emph{Andreotti}, 2004; \emph{Elbelrhiti et al.}, 2005; \emph{Claudin and Andreotti}, 2006]. In particular, this scaling was used to produce centimetric barchans at the scale of laboratory experiments [\emph{Hersen et al.}, 2002; \emph{Endo et al.}, 2004; \emph{Hersen}, 2005] and to understand the size of the dunes on Mars [\emph{Claudin and Andreotti}, 2006].

Sand was sampled in different places of the Moroccan dune field. The samples are from the sand right in the middle of the dune avalanche slip face. The corresponding PDFs (weighted in mass) are plotted in figure~\ref{ControlParams}b. They are well fitted by a log-normal law $P(d) \propto \exp \left [ - \left ( \frac{1}{\sigma} \ln \frac{d}{\la d \ra} \right  )^2 \right ]$. The fits give values for $\la d \ra$ between $165$ and $185~\mu$m with no particular trend. The values of the dimensionless width $\sigma$ of the distribution is of the order of $0.25$. This means that, in comparison to its value at the peak, the probability drops down by a factor of $2$ for a grain size either smaller or larger by a factor of $e^{0.25\ln\sqrt{2}}$ with respect to $\la d \ra$.

Therefore, apart from some coarse grains systematically visible at the toes of dunes in the form on mega-ripples patterns, the grain size distribution in the studied region is remarkably uniform with a typical size around $175~\mu$m. We can then conclude that, although some correlation of $W$ with $d$ is expected, grain size distribution is not the parameter responsible for the corridor pattern of this barchan field, and the size selection must have another cause.

As for the topography, the map of the region indicates that the plateau presents a slight slope in the cross wind direction which is estimated at most between $10$ and $20$~m for $5$~km ($0.3 \%$). However, it is unlikely to correlate such a regular slope with a sharp transition from small to large barchans. Finally, as we could observe in the field, the vegetation as well as the nature of the soil is very uniform everywhere.

%__________
\subsection{Wind velocity across the dune field}
Another important factor is of course the wind intensity. In particular, the only consistent theory for the saturation length $\ell_s$ [\emph{Andreotti}, 2004] (and thus the wavelength $\lambda_m$) predicts a sub-dominant variation with the wind velocity. On can wonder whether this could however influence the size selection of mature dunes.

We have measured the wind velocity along a road crossing the dune field in its northern part--~near Tarfaya, zone \textsf{A}. These measures were performed with a standard cup anemometer fixed at the top of a $3$~m high mast. We checked for each point that no neighboring dunes could influence the measurements. In order to remove the possible slow drift of the wind speed, a second fixed anemometer was used as a reference at the top of a dune -- as a general feature, the wind gently varies up in the morning to mid-day and then down in the afternoon. The measures are displayed in figure~\ref{ControlParams}c and show an almost perfectly flat profile \emph{versus} $y$. In particular, no sharp variation is visible around the location $y=0$ where the change from small to large dunes is observed. Moreover, no significant relief which could locally modify the trade wind flow is present. 

We have seen that the size and spacing in the barchan field of La Pampa de la Joya is very different from that of Morocco. One could imagine that this is to be related to the smaller value of $Q$ in Peru. However, the RDP or DP intensity factors mostly reflect the fraction of time during which there is transport i.e. during which the wind strength is above the transport threshold. Comparing the typical wind velocities in the two fields when sand transport is effective, we actually get very similar values. We thus conclude that the observed differences between the two fields do not result from different wind strengths. Once again, the origin of the corridor pattern is thus to be found from another parameter.

%___________________________________________________________________
\section{Dynamical mechanisms regulating the dune size}

We have described above the dune size distribution and the corridor structure in barchan fields. We have shown that the natural external parameters such as the wind velocity or the sand flux cannot explain such a pattern. We thus turn to the investigation of dynamical mechanisms regulating the dune size. We started to address these questions in previous papers -- see [\emph{Hersen et al.}, 2004; \emph{Elbelrhiti et al.}, 2005] -- and report here detailed field data to discuss the two main mechanisms. First, the volume of a dune evolves by sand supply on its stoss side and sand leak by its horns. Second, exchanges of mass occur due to interactions between dunes (collision and calving). These processes at hand, it is then interesting to trace their imprint in the dune landscape.

%__________
\subsection{Flux balance on a single dune}

\begin{figure*}[t!]
\includegraphics{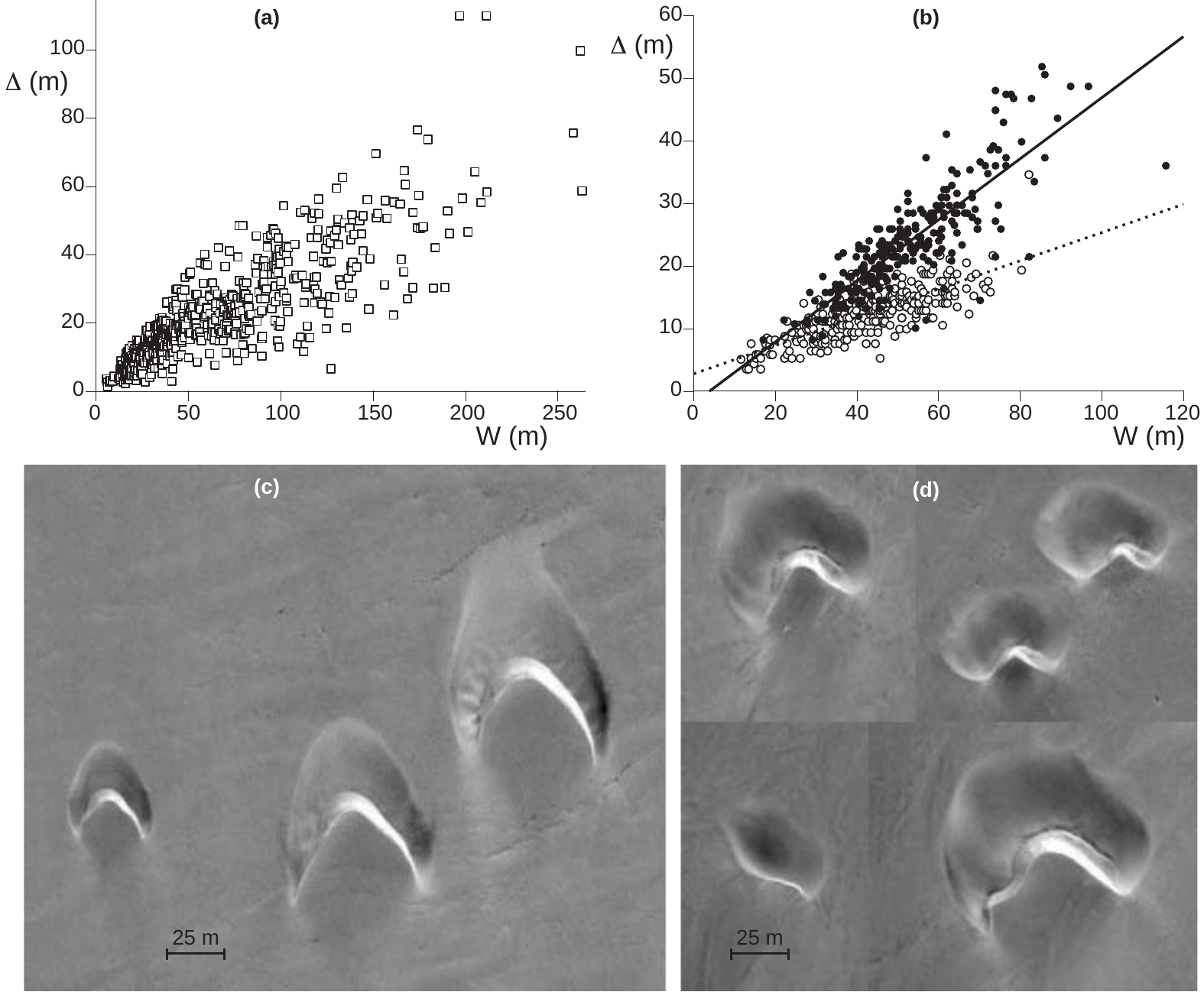}
\caption{(a) and (b) Total horn width $\Delta = \Delta_l + \Delta_r$ as a function of the dune width. In panel (a) the data come from zone \textsf{B} ({\small $\square$}). $\Delta$ is roughly proportional to $W$. Panel (b) is for La Joya, zones \textsf{E} ({\Large $\bullet$}) and \textsf{F} ({\Large $\circ$}). The photo of zone \textsf{E} is dated May 2003 and shows barchans with thick horns (panel (d)). Their blunt morphology with a small reversed slip face at the brink is the typical result of a strong storm blowing at an inclined direction with respect to the average wind. The photo of zone \textsf{F} is dated Feb. 2005 and the barchan shape (panel (c)) is that generated by the average wind. The two data sets can clearly be distinguished (solid and dotted lines in (b)).}
\label{DeltaofW}
\end{figure*}

The amount of sand that a dune of width $W$ receives per unit time when submitted to a flux $q$ is easy to compute: a simple cross section argument leads to $\phi_{\rm in} = q W$. The estimation of $\phi_{\rm out}$ requires more caution. As barchans loose sand by their horns, we can write $\phi_{\rm out} \sim Q \Delta$, where $\Delta$ is the total horn width of the considered dune, as previously discussed above for the free flux, see equation~(\ref{qfree}). We therefore extracted from our data $\Delta$ as a function of $W$. The point is that, if barchans are in an equilibrium state with a well defined width for a given input flux, we expect large dunes to loose proportionally more sand than small ones. In a previous numerical study [\emph{Hersen et al.}, 2004], we have exhibited the unstable nature of isolated dunes submitted to a strictly unidirectional wind, and which lead to ever growing or shrinking barchans.

Data from zone \textsf{B} are plotted in figure~\ref{DeltaofW}a. They are quite scattered, but to first order, they satisfy  $\Delta \propto W$ so that the balance equation $\phi_{\rm in} =\phi_{\rm out}$ which would hold for a steady dune cannot select a unique width. Note that in our numerical work [\emph{Hersen et al.}, 2004], the curve $\Delta$ (or $\phi_{\rm out}$) against $W$ was obtained with dunes artificially maintained in equilibrium by means of semi-periodic boundary conditions. By contrast, the data shown in figure~\ref{DeltaofW} corresponds to the same measurement performed on individually unstable dunes inside a statistically steady field.

The data from La Joya field shed an interesting light on this problem. The photos of zones \textsf{E} and \textsf{F} dates respectively back to May 2003 and Feb. 2005. They show barchans of distinctly different morphologies. In the later case, the dunes have their standard crescentic shape with slender horn tips (figure~\ref{DeltaofW}c), whereas in the former one the barchan present a blunt outline with thick flanks (figure~\ref{DeltaofW}d). The dunes of zone \textsf{E} furthermore display a small reversed slipface at their brink. Having directly observed this phenomenon many times in the field, we can ascribe it to a storm blowing from an anusual direction. Moreover, the collisions can be forgotten in La Pampa de la Joya as the barchans density is very low. The two data sets for $\Delta$ \emph{vs} $W$ in zones \textsf{E} and \textsf{F} can clearly be distinguished in the plot of figure (figure~\ref{DeltaofW}b): $\Delta$ for the smaller dunes does not differ much from a set to the other, whereas larger dunes can typically double their horn width after the storm. As mentioned in [\emph{Elbelrhiti et al.}, 2005], this size dependence in the sand loss can be interpreted with the induction of surface waves on the horns and flanks of dunes whose size is larger than several times $\lambda_m$. The response of dunes to the variations in wind direction (in particular storms) leading to an extra-leak of large dunes in the end controls the regulation of the size of these quasi-isolated barchans of La Joya.

%__________
\subsection{Mass exchange through collision}

\begin{figure*}[p]
\includegraphics{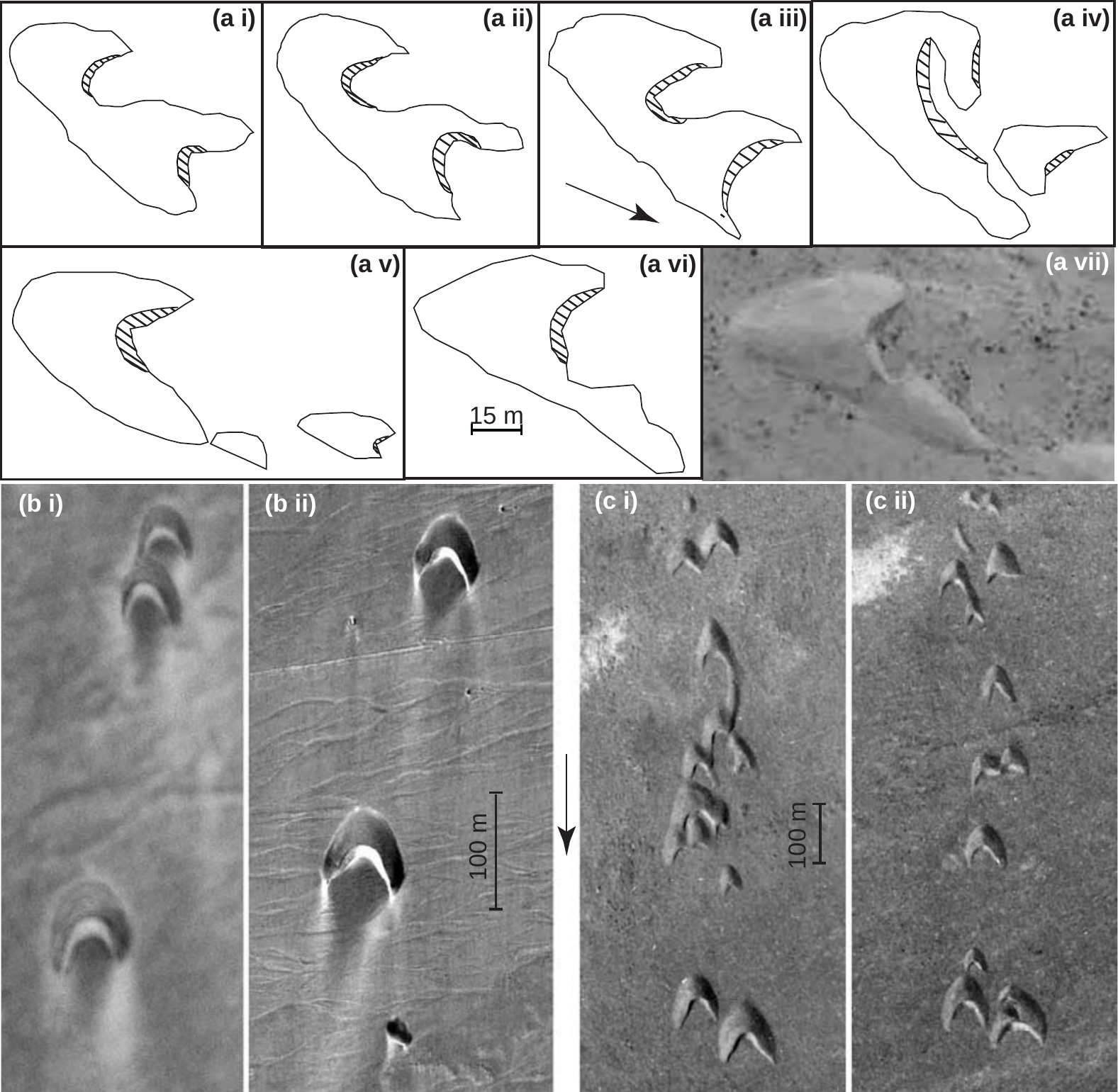}
\caption{Dynamical evolution of dunes of various sizes. (a) Detailed contours of two interacting small barchans in zone \textsf{A}. Dates are 27$^{\rm th}$ Jul. 2003, 23$^{\rm rd}$ Aug. 2003, 16$^{\rm th}$ Jan. 2004, 18$^{\rm th}$ Apr. 2004, 2$^{\rm nd}$ Aug. 2004 and 4$^{\rm th}$ Jun. 2005 for panels (a i) to (a vi) respectively. The photo of panel (a vii) dated May 2005. (b) The collision of the two dunes at the top resulted into a partial fusion. The time interval between the two photos is $47$~years. They are from La Joya, zone \textsf{F}. (c) The central area of photo (c i) has been subjected to significant rearrangements. In particular, the large barchan has significantly lost weight. The time interval between the two photos is $4$~years. They are from Atlantic Sahara, zone \textsf{D}. Arrows indicate wind direction.}
\label{Collisions}
\end{figure*}
\begin{figure}[t]
\includegraphics{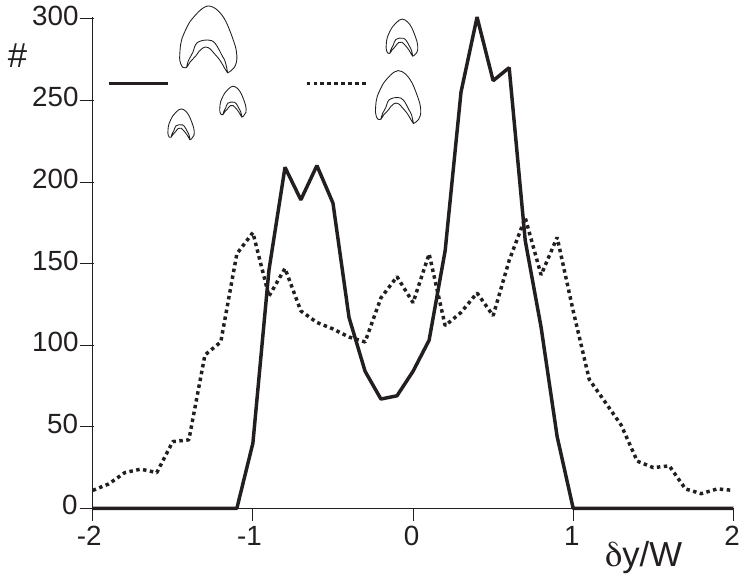}
\caption{Histograms of the rescaled transverse distances $\delta y / W$ between a dune and its next downwind neighbor (see figure~\ref{schemafluxes} for the definition of $\delta y$). As schematized in inset, the solid (dotted) line is for the case of a larger (smaller) upwind dune. These data come from the Moroccan fields. Although more scattered due to less statistics, data from La Joya field show similar curve shapes.}
\label{canard}
\end{figure}

In Morocco, the situation is different: the dune density can be high and collisions are very frequent. As illustrated in [\emph{Elbelrhiti et al.}, 2005], only the largest dunes can be identified after a time period of $30$ years and collisions play a major role in the sand transfers between the dunes. We have followed in detail the time evolution of two interacting dunes in zone \textsf{A}. The different contours are displayed in figure~\ref{Collisions}a. One can see that the dynamics is complex, involving a partial fusion together with a few ejections. By the use of the volume calibration law (\ref{equaVofW}), we can quantitatively estimate the mass balance before and after the collision. In panel (a i), the upwind smaller dune contains $600$~m$^3$ of sand, whereas the volume of the larger one is $800$~m$^3$. In panel (a vi), the resulting dune has grown up to $1400$~m$^3$. At least $150$~m$^3$ have been lost in the form of four barchans calved at elementary size -- typically $W \sim \lambda_m/2 \sim 10$~m, see e.g. panels (a iv) and (a v). There has thus been an additional supply during this period, as proved by the comparison between (a i) and (a iii). 

We have identified few potentially similar events in the Peruvian field such as that shown in figure~\ref{Collisions}b -- in this case intermediate photos are unfortunately not available. Such an interaction is typical for rather small dunes (few times $\lambda_m$ in size) and is also well reproduced in subaqueous experiments [\emph{Endo et al.}, 2004; \emph{Hersen}, 2005; \emph{Hersen and Douady}, 2005]. Examples of collisions involving large dunes are discussed in [\emph{Elbelrhiti et al.}, 2005]. They typically result in an intense calving in the wake of the barchan horns, leading to a significant size reduction of the target dune. In figure~\ref{Collisions}c is displayed an important rearrangement in zone \textsf{D} in a four years interval. The central large dune get destabilized and ejects several smaller ones. Some fusions also occurred upwind. In the end, these events effectively result in a higher sand leak for the larger barchans, preventing them from growing indefinitely which is what we would expect with pure fusions.

In order to investigate the importance of this barchan ejection process at the horns of large dunes, we studied how correlated the position of a dune with its next downwind neighbor is. As a matter of fact, all the calved dunes that we have been following in the field for more than five years (more than one hundred) have remained aligned with the horn of their mother dune. Conversely, the alignment of a small dune with the horn of a large one could be systematically associated with a former surface wave. More precisely, we have computed the histogram of the lateral distance $\delta y$ between the centers of such two dunes (see figure~\ref{schemafluxes}). In order to compare dunes of different sizes, this distance is rescaled by the width of the upwind barchan. As small dunes get ejected from large ones and not vice-versa, it makes sense to differentiate the case where the upwind dune is larger or smaller than its downwind neighbor. In figure~\ref{canard} we plot the two corresponding histograms. The former case shows a double peaked curve confined between $\pm 1$ in rescaled units, whereas the latter is broader with a flat profile. This double peak is the signature of the  frequently observed `duck fly'  packing arrangement where backs and horns of neighboring dunes almost touch each other. It demonstrates the ubiquity of the calving process, and thus the regulation of masses by dunes interactions.

%__________
\subsection{A dune landscape reading}

\begin{figure*}[p]
\includegraphics{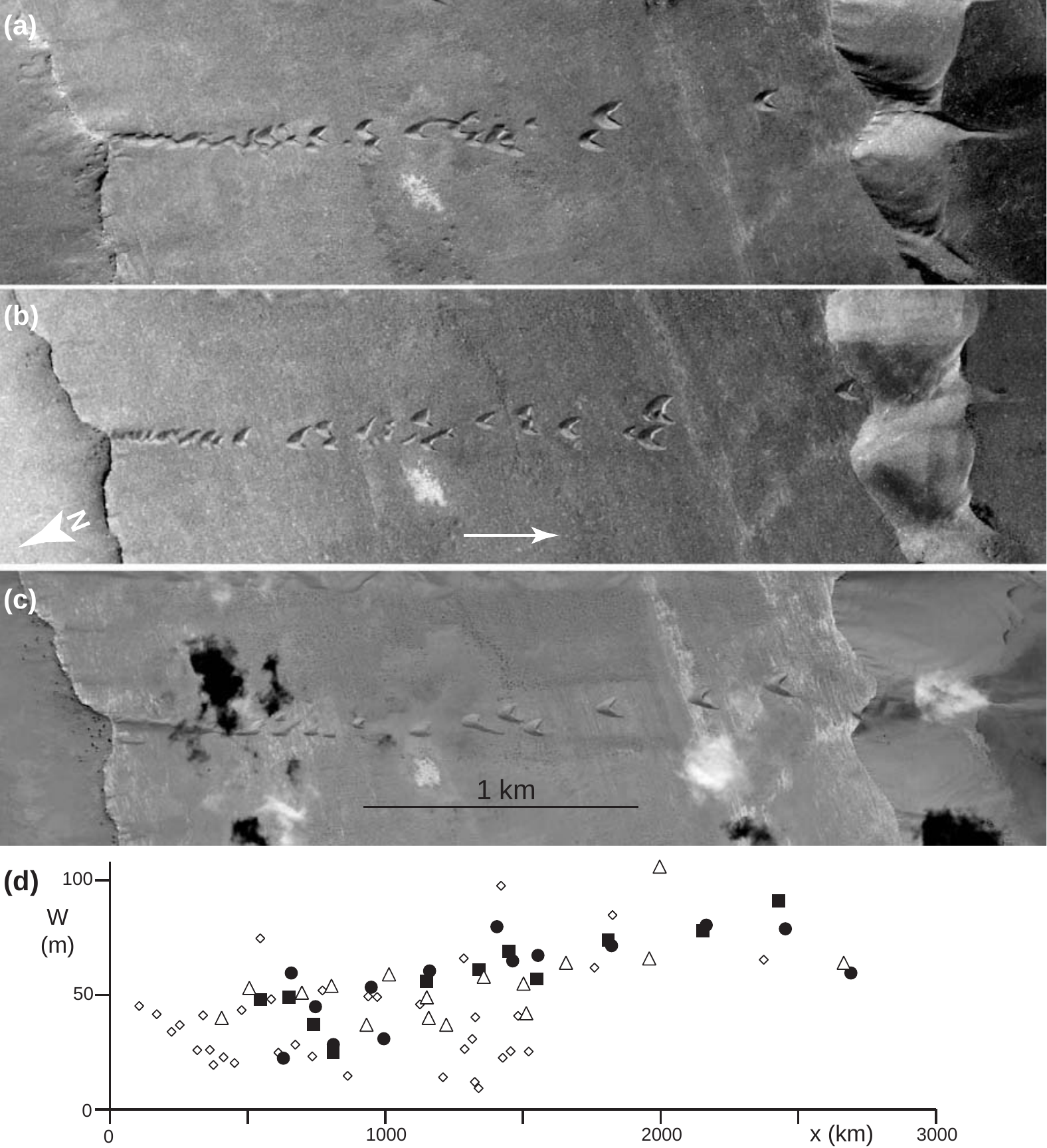}
\caption{Tizfourine `mono-corridor' (zone \textsf{D}). Photos dated 1975 (a), 1979 (b) and May 2005 (c). (d) Dune width plotted against its downwind distance from the northern cliff. $\lozenge$, {\small $\triangle$} and {\small $\blacksquare$} symbols are for the photos in (a), (b) and (c) respectively. The data plotted with {\Large $\bullet$} come from our GPS contours dated Aug. 2004.}
\label{Monocouloir}
\end{figure*}
\begin{figure*}[p]
\includegraphics{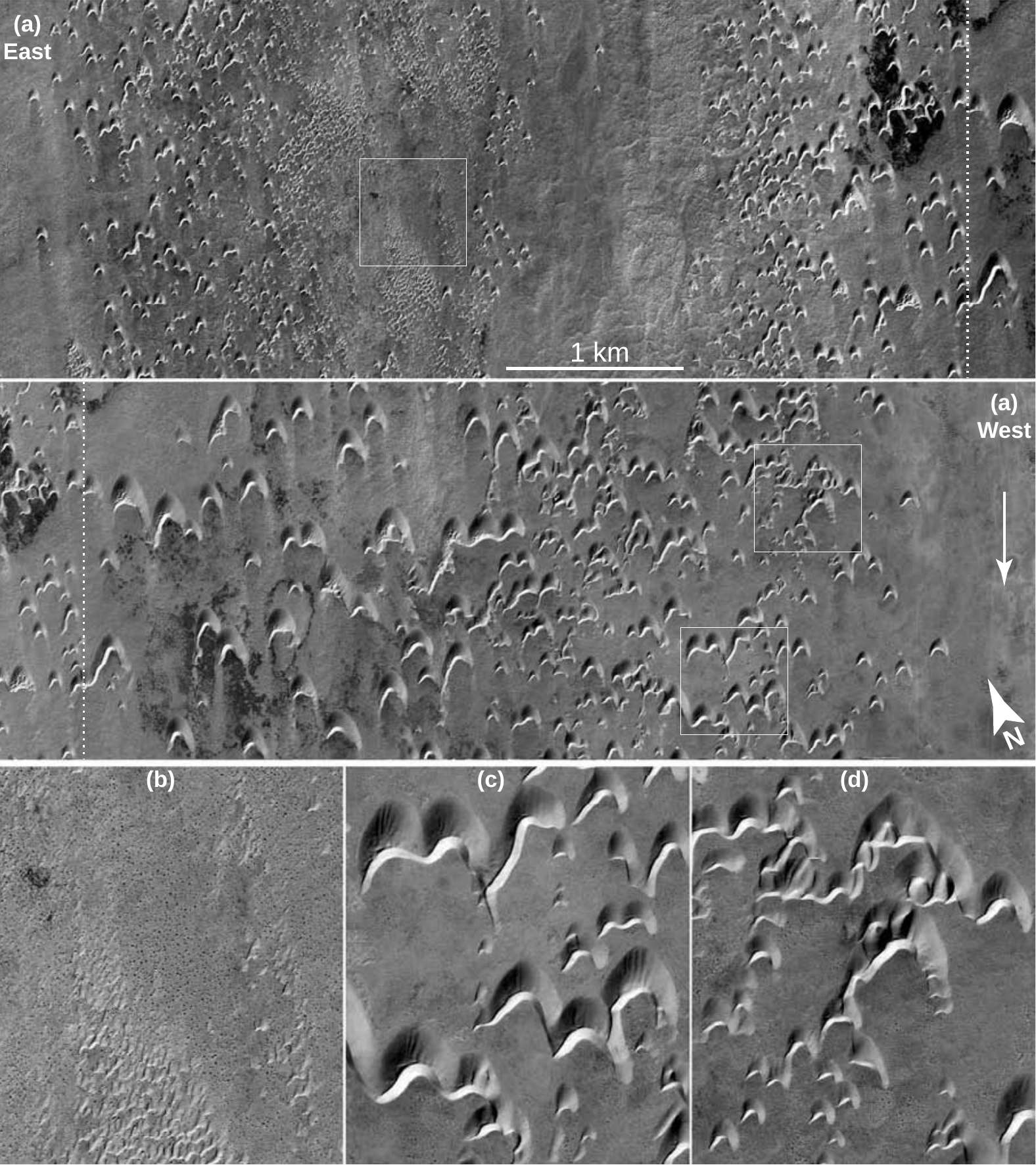}
\caption{Eastern (top) and western (below) parts of a cross wind stripe of the barchan field in zone \textsf{C} from which three places have been rxpanded -- note the overlap of the two photos indicated by the dash line. Zoom (b): a hole in the small-dune corridor and renucleation of the barchans further downwind. Zoom (c): surface instability in the form of undulations on the back and flanks of the dunes, induced by the storms displayed in figure~\ref{storms}. Zoom (d): calving processes leading to `duck fly' spatial organization. Note also the breaking of the largest barchan's left horn. This photograph dates back to the 5$^{\rm th}$ Jan. 2005.}
\label{zooms}
\end{figure*}

Having now analyzed the two principal dynamical processes at work in a barchan field, it is interesting to turn back to a reading of the dune landscape, looking for traces of what we have discussed above. A first place to examine is a small field located on a plateau between two depressions (zone \textsf{D} on the map displayed in figure~\ref{3zones}a). The topography is such that a point-like source of sand at the top of the cliff generates a single channel of barchans, see figure~\ref{Monocouloir}. As can be seen on the photos, this peculiar field starts in fact with a kind of elongated tongue of sand from which small dunes are calved. After detachment, the barchans propagate downwind while growing in size, and eventually disappear over the downwind cliff. More quantitatively, the dune width $W$ is plotted against the downwind position $x$ in panel (d) of figure~\ref{Monocouloir}. Although scattered, there is almost a factor of $2$ between size of the dunes at the beginning of the channel and those at the end. This coarsening is a persistent phenomenon as it is here observed in a coherent way over a period of $30$ years, which is long enough to renew most of the dunes in this small field. Following the above analysis, this is the signature of numerous collisions processes generating partial fusions.

\begin{figure*}[p]
\includegraphics{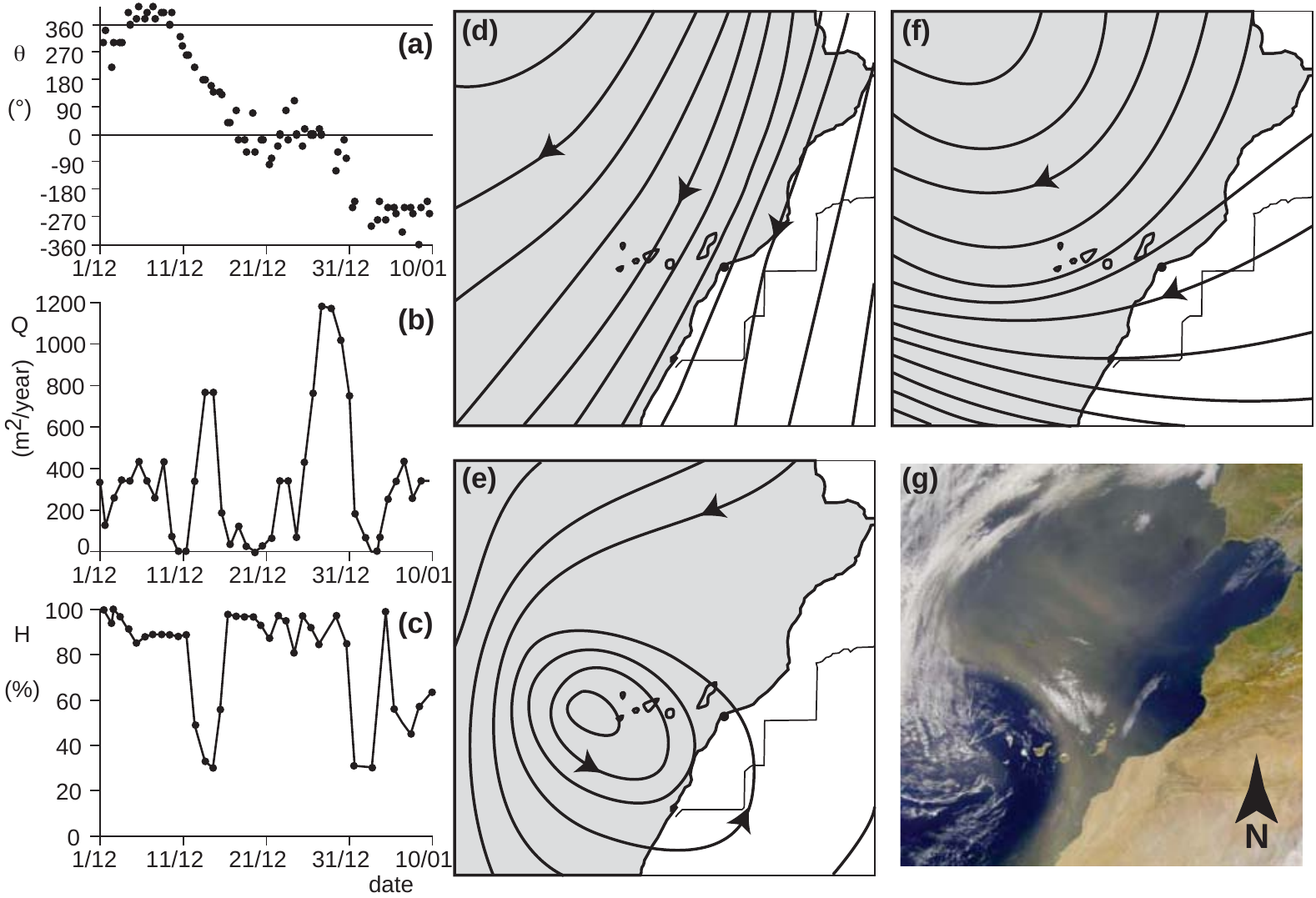}
\caption{(a-c) Meteorological data measured in Tan-Tan between the 1$^{\rm st}$ Dec. 2004 and the 10$^{\rm th}$ Jan. 2005. During this period, the dunes have been submitted to two storms, characterised by a strong dry wind coming from inland. The aerial picture of figure~\ref{zooms} has been taken on the 5$^{\rm th}$ Jan. 2005.  (a) Wind direction. (b) Reference saturated flux $Q$ on solid ground. (c) Humidity. (d-f) Wind streamlines reconstructed from QuickSCAT measurements and the weather stations of the region. The wind speed is inversely proportional to the distance between the streamlines. (d) Wind measured during a typical trade wind period.  The Azores anticyclone is located North-West of the map. (e) Wind on the 14$^{\rm th}$ Dec. 2004. The Azores anticyclone has moved over Europe, allowing the appearance of a cyclone inducing a dust jet over the ocean. (f) Wind on the 2$^{\rm nd}$ Jan. 2005. (g) Dust storm over Canarias Islands (credits: Nasa, \texttt{http://visibleearth.nasa.gov/}).}
\label{storms}
\end{figure*}

Other illustrative configurations can be found with a crosswise scan in the large field of zone \textsf{C}. This is the purpose of figure~\ref{zooms} where three places of interest have been expanded. Panel (b) shows a `hole' in the field, in the sense that this place is locally empty of dunes. The dunes however reappear further downwind in a dense packing of elementary barchans. Although the precise reason of this local disappearance of dunes is unknown, this phenomenon is not that surprising reminding that the total flux is only of the order of half of $Q$ (figure~\ref{FreeBulkTot}). In panel (d), we recognize calving processes and corresponding typical spatial packing arrangements as well as numerous sub-structures on the flanks of the large barchans which are the sign of strong interactions between the dunes. In other places (panel (c)), the dunes show typical surface undulations that we can associate to strong oblique winds. As a matter of fact, the analysis of the meteorological data of the period preceeding the date at which the photograph of figure~\ref{FreeBulkTot} has been taken shows that two storms had recently occured. The temporal recordings of the reference flux (modulus and direction) and the air humidity,  displayed in figure~\ref{storms}a-c, allow to date these events around the 14$^{\rm th}$ Dec. 2004 and the 2$^{\rm nd}$ Jan. 2005. Such storms usually correspond to a situation where the Azores anticyclone generating the trade winds (figure~\ref{storms}d) moves to Europe, which allows for the appearance of a cyclone further south (figure~\ref{storms}e-g). As a result, strong and dry winds, called Chergui, blow from inland over the Atlantic Sahara. These storms mostly occur during winter time, at a typical frequency of ten events per year. They induce an important transverse transport --~$Q$ is typically ten times larger than its averaged value~-- during few days. This is sufficient to induce lasting disturbances on the shape of the dunes, without, however, moving significantly their centre of mass. 

In summary, the whole field, although well organised in corridors, should not be thought of in a static -- or rather purely kinematic -- manner with an external driving, but as the place of numerous dynamical events, where small dunes undergo partial fusions and large ones calving processes in response to collisions due to faster small barchans running into them. In the `diluted field' case of La Joya where collisions are too rare to be a dominant process of size regulation, storms must suffice to prevent to the growth of the dunes by a temporary increase of their horn width and thus of the sand loss.

%___________________________________________________________________
\section{Conclusion}

%__________
\subsection{Summary of the results}
The main points of this paper can be summarized as followed:\\
$\bullet$ Two sites have been studied: the barchan field between Tarfaya and Laayoune (Atlantic Sahara) and that of La Joya (Peru). From a comparison of dune positions over time we obtained a relationship between their propagation velocity $c$ and their width $W$. Similarly, the use of ground photos led to get the dune volume $V$ as a function of $W$. Both of these empirical laws $c(W)$ and $V(W)$ are well defined over the whole range of dune size -- from $W \sim 20$ to $W \sim 600$~m.\\
$\bullet$ We performed a quantitative analysis of aerial photographs, looking in particular to the probability distribution function of the dune size, the average dune width $\la W \ra$ and the dune packing density $\eta$. A clear corridor pattern is revealed with a sharp variation in the transverse ($y$) direction of $\la W \ra$ and $\eta$, the smaller barchans being more densely packed. By contrast, these quantities are homogeneous in the windward ($x$) direction.\\
$\bullet$ Investigating the possible external parameter controlling the dune size in these fields, we measured the grain size distribution in different places of the field as well as the wind profile across the corridors. We also examined the sand flux profiles. Following \emph{Lettau and Lettau} [1969] we split it into the `bulk' flux corresponding to the amount of sand transported by the dunes themselves and the `free' flux for the sand transported in saltation between the dunes. We showed that none of these profiles presents a significant and systematic variation at the location of corridor change. Furthermore, as the quantitative value of the total sand flux is less (typically one half) than the saturated flux, it indicates the possibility to get under the same conditions a lasting sand transport with dunes or without i.e. in saltation over the solid ground only.\\
$\bullet$ We show that the dune size is controlled by two dynamical processes. First, changes in wind direction, and storms in particular, induce a broadening of large-dune horns. Second, the collisions between dunes strongly depend on their sizes. While small colliding dunes (whose rescaled width $W/\lambda_m$ is of the order of unity) exhibit a partial fusion, large ones (in terms of $W/\lambda_m$) get destabilized and generate a wake of small barchans of elementary size.

%__________
\subsection{Discussion}

Although the origin of the corridor structure is finally left unexplained, some key dynamical mechanisms have been identified that provide a new way to read and interpret the dune landscape. Moreover, our conclusions from field observations allow to reject the most obvious hypotheses concerning the selection of barchan size. The first idea would be that the dunes are individually stable so that they would essentially keep the size at which they were formed. A refined option is to consider that isolated dunes are unstable with respect to sand exchanges but that dune fields are made collectively stable by collisions, assuming that the collision of a small dune with a large one always leads to two medium sized dunes. Then, it can be inferred that after few collisions, all the dunes would have almost the same volume. Again, the average size would be determined by the entrance conditions i.e. by the distribution $P(w)$ at their place of birth. However, in the Atlantic Sahara, the dunes form roughly with the same size, whether that is on the beach or after a sebkha, but evolve towards very distinct states in the different corridors. Moreover, we have tracked several collisions after which the impacted barchan becomes larger. This evidences that the size selection in barchan fields does not result from collisions only and does not proceed from a memory of the conditions of formation.

Our results show that the mechanisms responsible for the size regulation are not scale invariant. For instance, the collision between a small and a medium dune does not resemble that of a medium and a large dune, since the length scale at which dunes form always remains prevalent in the process. We have also shown that the distributions of size are not similar in small (single peaked) and large dune (double peaked) corridors. In the later, the second peak corresponds to the dunes calved from the horns of the large ones.

Lastly, we have shown that the intermittency of the wind regime could play an important role in the selection. We have provided direct evidences that a storm i.e. a strong oblique wind, can increase significantly the dune output flux. In La Joya, the barchan field is so sparse that collisions hardly occur: the size control can be ascribed to the wind regime, only. In the Atlantic Sahara, the stormy winds mostly come from inland. This could result into a sand flux transverse to barchan corridors, breaking the symmetry between the edge and the bulk of the field, between its eastern and western parts. 

These conclusions point out several directions for future theoretical and field investigations. The models should be tested against the new data and possibly improved to correctly reproduce the dynamical response of barchans to perturbations such as storms and collisions. A key issue is to incorporate the wind regime into object oriented simulations, to see whether or not the corridor structure spontaneously emerge. Concerning the field work, a challenging issue is the direct and precise measure of the sand flux coming out of a barchan of given width. In relation to this question, a finer measurement of the horn width $\Delta(W)$ should be performed -- from aerial photos or in the field -- e.g. in dune configurations restricted to particular situations such as collisions.

\vspace*{0.5cm}

%___________________________________________________________________
\noindent
\rule[0.1cm]{3cm}{2pt}

This paper has benefited from  a careful reading and useful comments of A.B. Murray and R.C. Ewing. We wish to thank O. Duran, E. Parteli and H.J. Herrmann for discussions on dune size distributions. A.B. and P.C. acknowledge J. Olivieri institute and St Victor research center for hospitality. This study was supported by an `ACI Jeunes Chercheurs' of the french ministry of research.

%\newpage

\appendix

%___________________________________________________________________
\section{Field measurement techniques}
\label{AppA}

%__________
\subsection{DIY geographic information system}
Most of the measurements presented here are based on pictures and dune contours, measured with hand-carrying Garamin GPS receivers. The GPS resolution, checked at reference points, is of the order of $4$~m. Depending on the size of the dune, the GPS points are spaced by $5$~m to $20$~m. The aerial photographs are taken with a Leica Geosystems camera on $240$~mm $\times 240$~mm films, with a lens of focal length $88$~mm or $152$~mm. They are digitalized at $60$~pix/mm, which, depending on the altitude of the plane, gives finally a resolution between $20$~cm/pixel for a field of $2.7$~km and $70$~cm/pix for a field of $10$~km. The ground photographs are taken with a Leica digital camera ($2048$ x $1536$ pixels). The lens focal length and the size of the sensitive elements are calibrated in the lab.

\begin{figure*}[t!]
\includegraphics{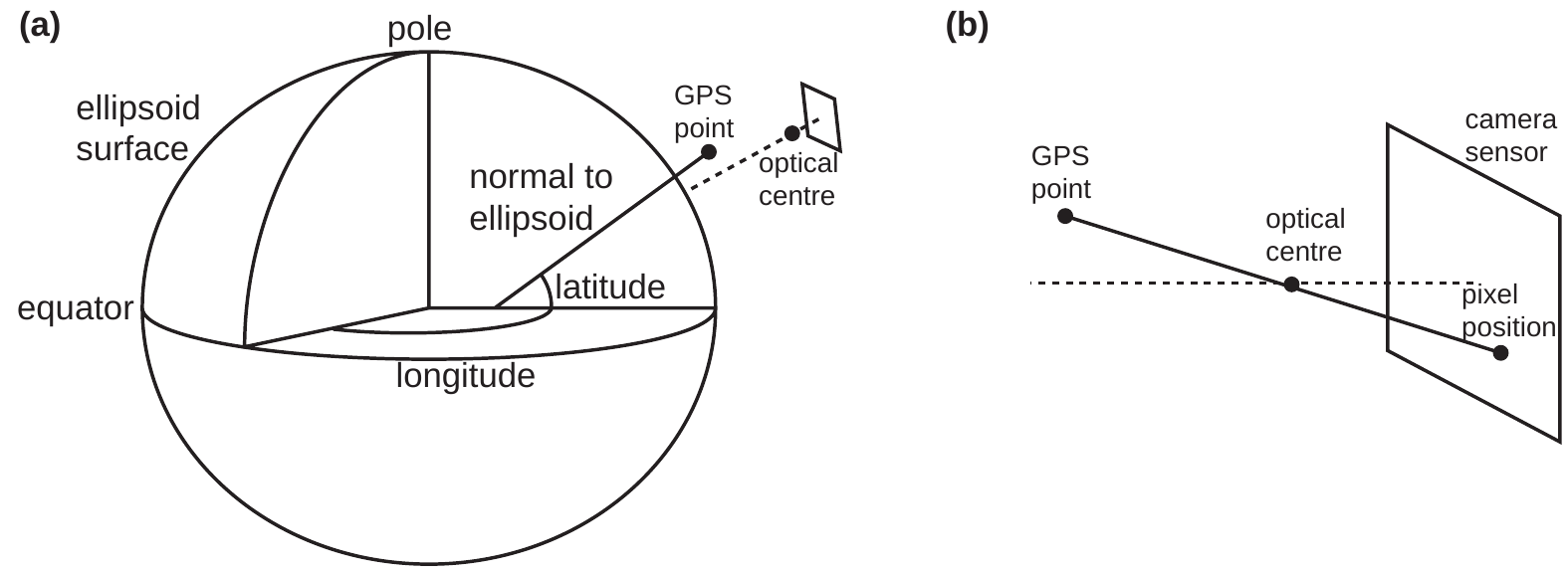}
\caption{Schematic showing (a) the convention for geodetic coordinates and (b) the projection of a point on a photograph.}
\label{DIYGIS}
\end{figure*}

Using a home made geographic information system, the GPS positions of the contours are superimposed on aerial or ground taken photographs. The principle of the projection used is shown in figure~\ref{DIYGIS}. The geodetic coordinates are converted into cartesian coordinates, using the WGS84 reference ellipsoid [see note]. The linear projection on the photograph is performed by introducing homogenous coordinates. The global position of the camera (latitude, longitude, altitude and the three Euler angles) is determined from the GPS positions of various landmarks (road, relief, cliff, track) visible on the photographs. This procedure ensures a global positioning of any visible point within a resolution of less than $4$~m. The relative resolution obtained for aerial pictures is typically metric but can be as small as centimetre for ground pictures of dunes.

\begin{figure*}[t!]
\includegraphics{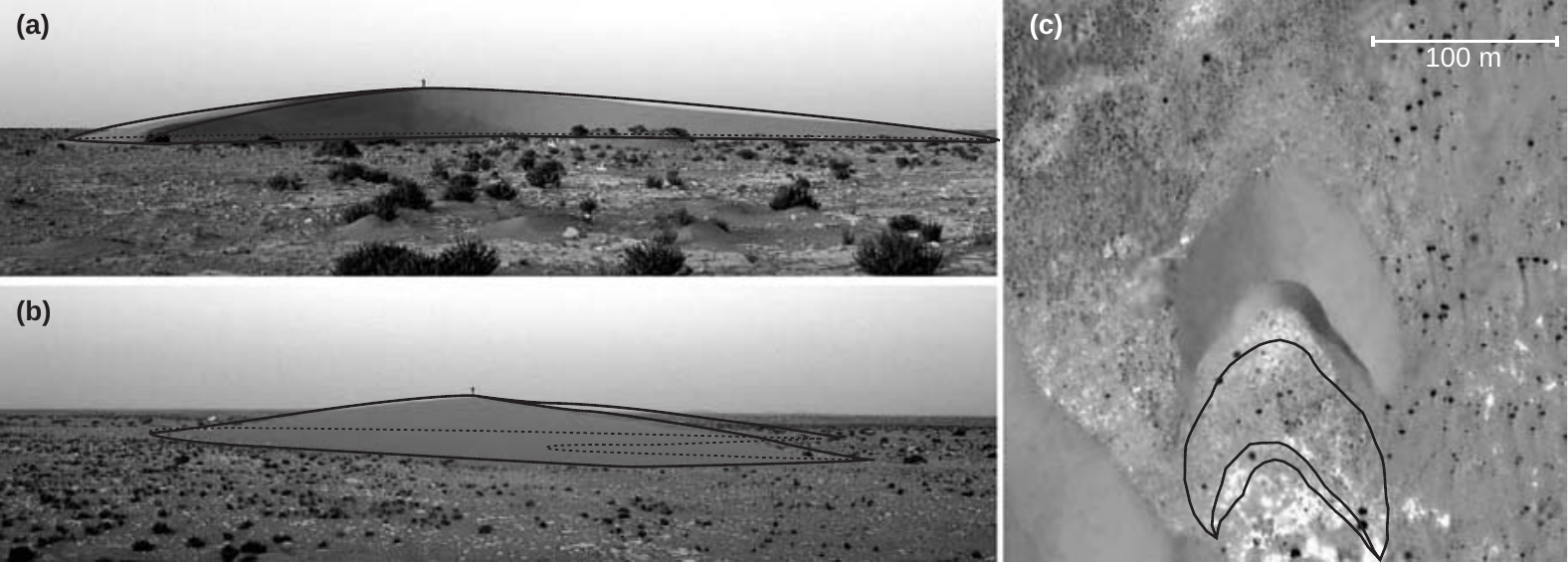}
\caption{Face (a), profile (b) and horizontal contour (c) of a dune -- `la grande blonde'. Its 3D geometrical reconstruction from which the dune volume can be computed is shown with the solid (or dashed when hidden) lines. In panel (c) these lines have been shifted downwind for comparison with the top view. In photos (a) and (b) a person at the top of the dune gives an idea of the vertical scale ($H \sim 8$~m).}
\label{FaceProfil}
\end{figure*}

The morphological characteristics of a dune (length $L$, width $W$, horn widths) are determined from either an aerial view or a GPS contour, by defining the six characteristic points shown in figure~\ref{6pts}. When the brink of the dune coincides with the crest, the height $H$ is determined from the measurement of the slip face length and of the avalanche slope. It can otherwise be measured using ground photographs (see figure~\ref{FaceProfil}). To determine the dune volume, we have developed a specific tool, which uses either three photographs (aerial, face and side photographs) or two photographs (face and side views) and the dune GPS contour. More precisely, we approximate the envelope of the dune using its contour and a point --~in general close to the brink~-- such that the profile along any cut of the dune passing through this point is a parabola. To reconstruct the avalanche slip-face, we use the property that the steepest direction of the slip face is perpendicular to the contour of the dune and that the slope is everywhere equal to the sand dynamic friction coefficient ($\tan 32^{\circ}$). Five parameters have to be tuned, looking simultaneously at the three views: the three dimensional position of the origin point and the slope of the surface at this point. The figure~\ref{FaceProfil} shows the quality of the reconstruction that can be achieved. Then, the volume $V$ of the dune can be easily deduced from this geometrical reconstruction.

%

%__________
\subsection{Morphology}
\begin{figure*}[t!]
\includegraphics{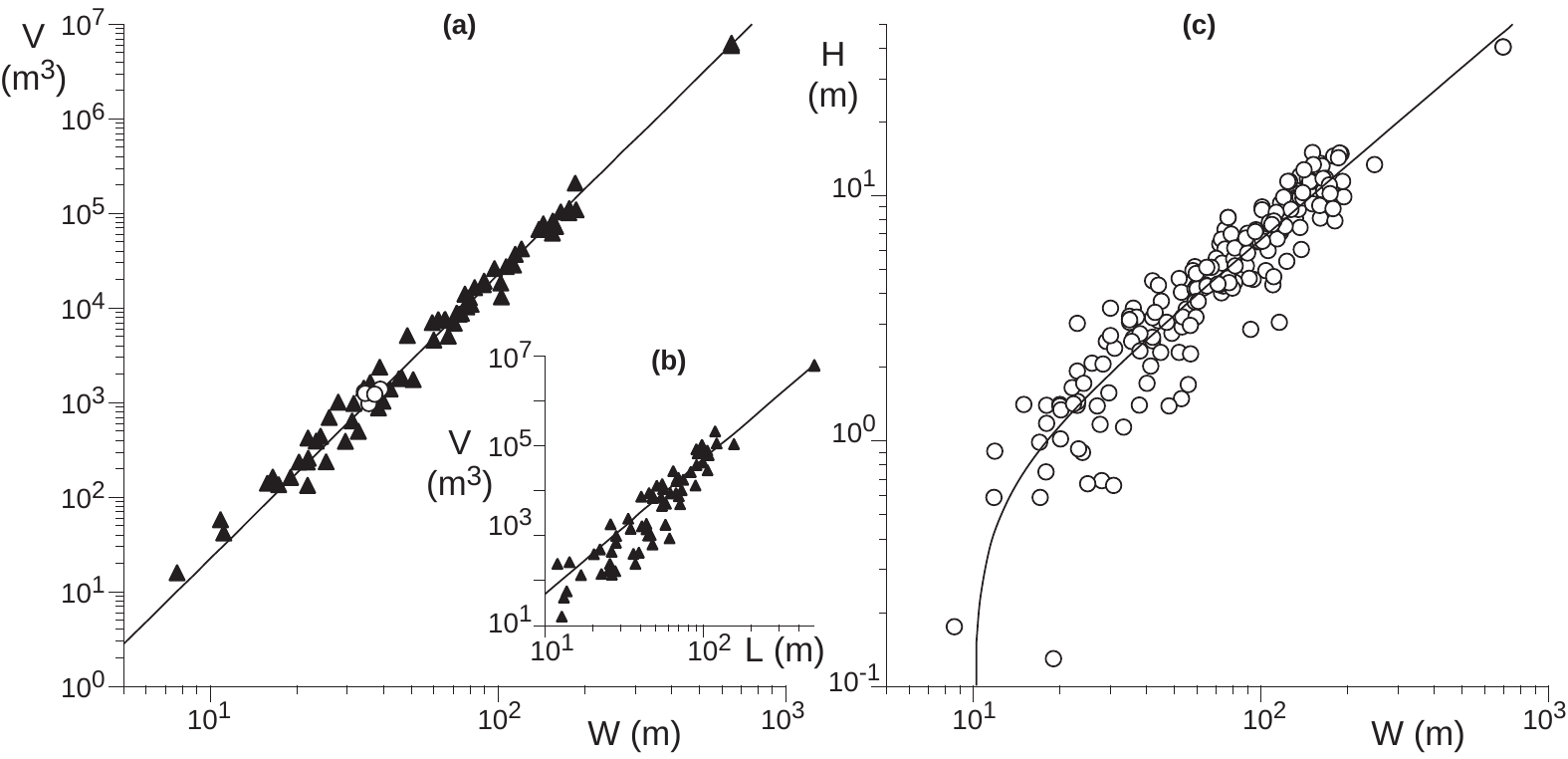}
\caption{(a) Dune volume as a function of its width. The solid line is the best power law fit $V \sim \frac{1}{40} W^3$. Note the four white circles around $W=30$~m which are the average measures of \emph{Lettau and Lettau} [1969] in La Joya barchan field. (b) When plotted against the dune length $L$, the data dispersion is larger. (c) Relationship between dune height and width. The solid line corresponds to an approximation of the data by the formula $H=\frac{1}{15}\sqrt{W^2-W_c^2}$, with $W_c \sim 10$~m.}
\label{figmorpho}
\end{figure*}

In agreement with other field data, reviewed for instance in [\emph{Andreotti et al.}, 2002a], the barchan dimensions (height $H$, length $L$ and width $W$) in the Moroccan field are linearly related to each other. An example of $H$ \emph{vs} $W$ is displayed in figure~\ref{figmorpho}c: dunes have a typical aspect ratio of $1/15$. Note that the linearity breaks down for small dunes as $H \to 0$ at a cut-off length $W_c \sim 10$~m, whose value is consistent with that of $\lambda_c$ discussed below in sub-section~\ref{dunesizedistri}. Note also the point at $H \sim 40$~m which corresponds to a mega barchan ($28^\circ 02'$N, $12^\circ 09'$W), and which is clearly scale separated from the other data points but still on the line.

As shown in figure~\ref{figmorpho}a, there is a satisfactory univocal relationship between $V$ and $W$ given by:
\begin{equation}
V \sim \frac{1}{40} W^3
\label{equaVofW}
\end{equation}
over two decades in width. Plotted against $L$ (figure~\ref{figmorpho}b), the volume scales as $V \sim \frac{1}{20} L^3$, but the data points are more dispersed. A slightly better collapse of the volume data (not shown) has been empirically obtained with $V \sim \frac{1}{38} L^{1/2} W^{5/2}$ but we checked that none of the conclusions reached here is affected by the choice of the formula for $V(W,L,H)$. This dispersion as well as that of the $H$ \emph{vs} $W$ plot is due to the fact that $L$ and $H$ are quantities sensitive to wind and sand flux changes. Indeed, we know from numerical simulations that the dune width is only evolving due to the small transverse flux, while the height and length can evolve more rapidly. As a result, instantaneous and fluctuating values of $H$ or $L$ are less representative of their average value than $W$ whose evolution time scale related to transverse grain motion is slower. As the dune width is an easy and robust quantity to measure of aerial photographs, such calibration curves let to compute all other barchan morphological characteristics from $W$.

\begin{figure*}[t!]
\includegraphics{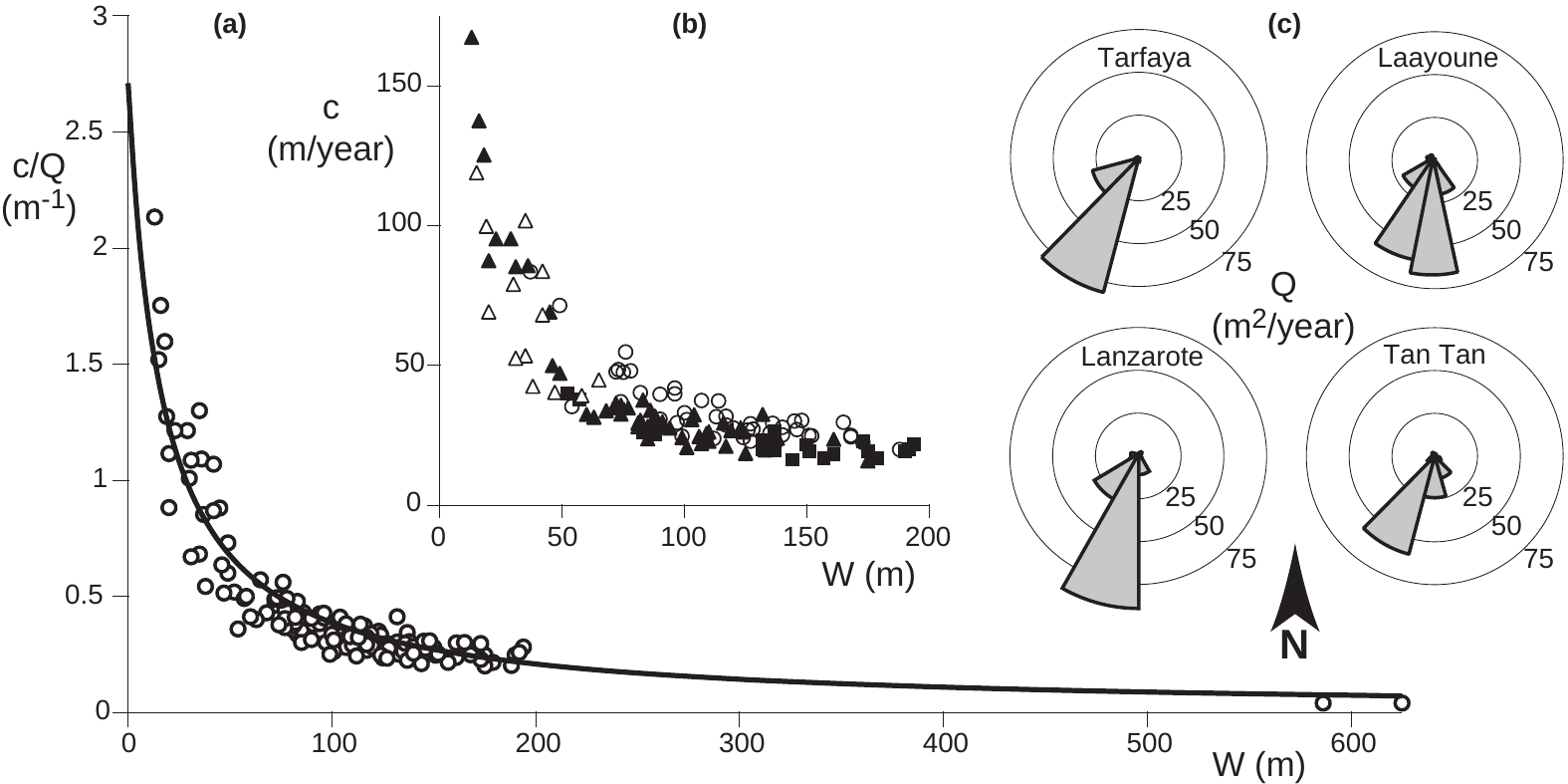}
\caption{(a) Propagation speed, rescaled by $Q$, as a function of $W$. The solid line is the fit of $c = \frac{bQ}{W+W_0}$, with $b \sim 45$ and which gives $W_0=16.6$~m. This (modified) Bagnold's law is valid up to the mega barchan points ($W \sim 600$~m). (b) Raw velocity data \emph{vs} $W$ measured at different places and for different time intervals: Tarfaya (zone \textsf{A}) Nov. 1979 -- Sep. 2004 {\large $\blacktriangle$}, Tarfaya (zone \textsf{A}) Mar. 31$^{\mbox{\scriptsize st}}$ 2003 -- Sep. 10$^{\mbox{\scriptsize th}}$ 2004 {\small $\triangle$}, Tah (zone \textsf{B}) Nov. 1975 -- Sep. 2004 {\small $\blacksquare$}, Laayoune (zone \textsf{C}) Aug. 1976 -- Sep. 2004 {\Large $\circ$}. (c) Sand flux roses computed with wind velocity time series from various places of the region of study: Tarfaya ($27^\circ55'$N, $12^\circ56'$W), Laayoune ($27^\circ09'$N, $13^\circ13'$W), Lanzarote ($28^\circ57'$N, $13^\circ36'$W) and Tan-Tan ($28^\circ27'$N, $11^\circ09'$W).}
\label{speedandroses}
\end{figure*}
%

%__________
\subsection{Propagation speed}
The displacement can be measured within a resolution of few meters by using either two GPS contours  (separated by typically one year), one GPS contour and an aerial photograph (with almost a $30$~years interval in Morocco) or two aerial photographs (separated by $47$~years in La Joya). This guarantees a precision that is less than a percent for the velocity $c$: as one can expect, the first method works fine for the smaller dunes, whereas the second one is better for larger ones as one need to be able to recognize the dune after several decades of evolution. $c$ and $H$ are related to each other by the conservation of matter, through the sand flux passing at the crest. Assuming that this flux is proportional to the saturated sand flux $Q$ over a flat bed i.e. the maximum amount of sand that can be transported by a wind of a given strength, Bagnold has suggested a law of the form $c(H) = aQ/H$, where the parameter $a$ is related to the speed-up factor ($\sim 1.4$) of the wind velocity on the back of a barchan. Here we express the velocity $c$ as a function of the width $W$ and we introduce a cut-off size $W_0$:
\begin{equation}
c = \frac{bQ}{W+W_0} \, ,
\label{equacofW}
\end{equation}
with $b \sim 45$. The existence of a cut-off length corresponds to the fact that very small dunes propagate at a finite velocity, as predicted by a linear stability analysis [\emph{Andreotti et al.}, 2002b; \emph{Elbelrhiti et al.}, 2005]. The fit of the data gives $W_0 \sim 16.6$~m which reasonably is in between $\lambda_c$ and $\lambda_m$.

To check this relation, $Q$ can be computed from wind speed time series. We obtain the time series for $Q$ using calibration curves provided by \emph{Iversen and Rasmussen} [1999], which, for $d=180~\mu$m, can be fitted by $Q=25 \frac{\rho_f}{\rho_s} \, \sqrt{\frac{d}{g}} (u_*^2 - u_{th}^2)$ with a good approximation [\emph{Andreotti}, 2004]. $Q$ vanishes below the threshold shear velocity $u_{th} \sim 0.1 \sqrt{\frac{\rho_s}{\rho_f}gd}$ with $\rho_s = 2650$~kg/m$^3$ and $\rho_f = 1.2$~kg/m$^3$. Then, the sand flux rose is computed by integration over one year of the saturated flux for each compass direction (30$^\circ$ wide bins). The sand flux rose is a standard tool to characterise the wind regime. An index of the directional variability of the wind is given by the ratio of the resultant drift potential (RDP) to the drift potential (DP). The DP is the integral of $|\vec{Q}(t)|$ whereas RDP is the modulus of the integral of $\vec{Q}(t)$. We obtain a RDP/DP of $0.91$ (slightly less, $0.89$, for the flux/velocity relationship commonly used), which corresponds to a narrow unimodal wind regime in the classification proposed by Fryberg and Dean. The region around Tarfaya (Atlantic Sahara) is thus comparable to other regions on Earth where the wind regime is very unidirectional (RDP/DP of $0.87$ in Walis Bay, Namibia, $0.91$ in Chimbote, Peru, $0.92$ in Bulgan, Mongolia, and $0.97$ in Aranau, Brazil). 

If we average expression (\ref{equacofW}), we see that the mean velocity is related to the mean flux (the RDP value), noted also $Q$ for simplicity. Using three airport wind time series (Laayoune, Lanzarote and Tan-Tan), and one obtained in the middle of the dune field, close to Tarfaya, we obtain values for $Q$ between $60$ and $90$~m$^2/$year. The mean dune velocity also varies from year to year and from place to place. As a matter of fact, the raw velocity data shown in panel (b) of the figure are quite dispersed. In order to improve the data collapse, we divided each data set by the corresponding value of $Q$ obtained from a fit of the considered set. These values for $Q$ read, all in m$^2/$year, $84$ ({\large $\blacktriangle$}), $75$ ({\small $\triangle$}), $67$ ({\small $\blacksquare$}) and $90$ ({\Large $\circ$}), see figure~\ref{speedandroses} for the meaning of the symbols.

\begin{figure}[t!]
\includegraphics{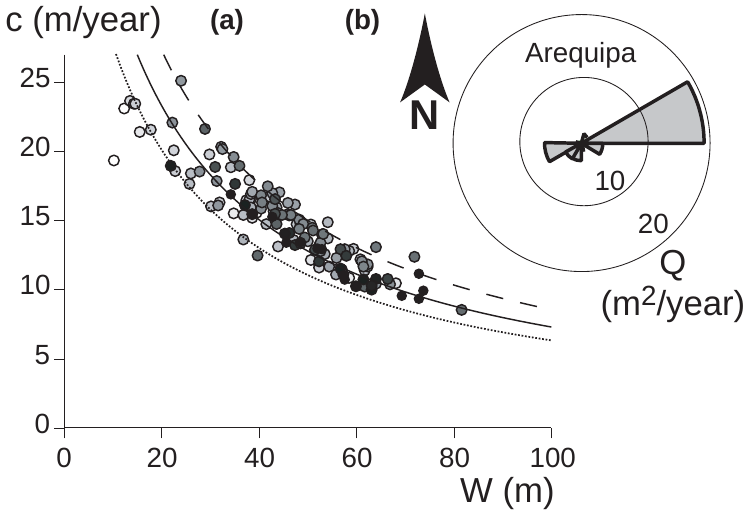}
\caption{(a) Dune velocity \emph{versus} $W$ measured in zone \textsf{F} over a time interval of $47$ years. The different symbols are for different parts of the field (the darker, the further east). The solid lines are data fits with equation (\ref{equacofW}): dotted line eastern and western parts, dashed line central part, solid line the whole field. The fits give $Q \sim 20$~m$^2$/year. (b) Sand flux rose at Arequipa ($16^\circ24'$S $71^\circ31'$W). As evidenced on the map of figure~\ref{JoyaMap}, the wind direction has turned by almost $90$ degrees in comparison with La Joya, but the wind regime remains mono-directional.}
\label{c_W_rose_Joya}
\end{figure}

Figure~\ref{c_W_rose_Joya} shows the displacement of dunes in La Pampa de la Joya from Oct. 1958 to Fev. 2005. Due to a rather low dune density, the number of collisions or major rearrangement events was so small that most of the dunes could be traced over this very long period of time. In agreement with what was mentioned by Lettau and Lettau, some systematic variations of the velocities across the field are observed (see figure~\ref{c_W_rose_Joya}). Another particularity of the wind flow of the region is that, although a unimodal regime is well kept, the wind main direction continuously rotates from south to north, due to local relief (mountains). As a result, the sand flux rose computed with data from Arequipa (no wind data time series available at La Joya unfortunately) shows a resultant direction almost perpendicular to that of La Joya (see figure~\ref{c_W_rose_Joya}b). With the graph of figure~\ref{speedandroses}, our results demonstrate the validity of Bagnold's law over a wide range with all wind speed influence captured in the value of $Q$.

%___________________________________________________________________
\section{Dune size distribution}
\label{AppB}
\begin{figure}[t]
\includegraphics{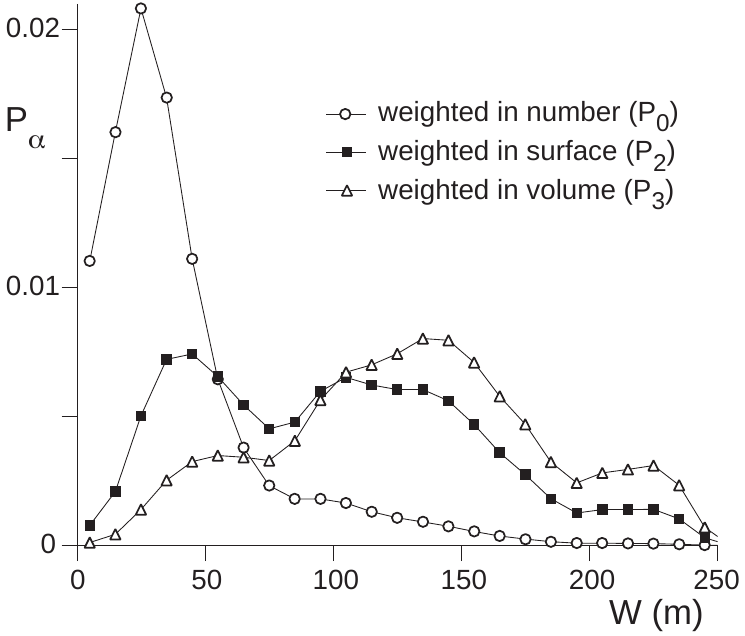}
\caption{Probability density functions $P_\alpha(W)$ computed over the Laayoune field (zone \textsf{C}). $\alpha=0$ (weighted in number) favors small dunes whereas $\alpha=3$ (weighted in volume) increases the importance of large ones. $\alpha=2$ (weighted in surface) gives a good neutral compromise. }
\label{PDFSize}
\end{figure}

Several options are available to define a dune size distribution. One can simply count the number of dunes $\delta N(W,\mathcal{A})$ in that domain whose width is between $W$ and $W+\delta W$. The PDF is then computed as $P(W,\mathcal{A})=\delta N(W,\mathcal{A})/\delta W/N_\mathcal{A}$, where $N_\mathcal{A}$ is the total number of dunes in the domain. This is valid in the regime where the shape of the PDF is insensitive to the value chosen for the interval $\delta W$. Another possibility is to weight the value of $P$ with some supplementary factor, that is a function of the dune size, e.g. the width to some power $\alpha$. A fairly more general formula is then
\begin{equation}
P_\alpha (W,\mathcal{A}) =\frac{\delta N(W,\mathcal{A}) / \delta W \, W^\alpha}{\sum_{i \in \mathcal{A}} W_i^\alpha} \, ,
\label{PofW}
\end{equation}
where the integer $i$ runs over all dunes in the domain $\mathcal{A}$. The standard $\alpha=0$ case gives favor to small dunes, whereas a larger value of $\alpha$ increases the weight of large ones. This analysis is important for what we want to call a `corridor' of barchans or how we can classify a group of dunes. Take for example a domain where some large dunes are surrounded by many little ones. Should it be effectively thought as a bunch of small dunes because the number of large ones is only small fraction of the total number of dunes, or should it be considered as a loosely group of large dunes because they are the ones which contain and carry most of the sand, the contribution of the small ones being negligible? In figure~\ref{PDFSize} we plot $P_\alpha(W)$ for different values of $\alpha$, computed over the whole zone \textsf{C} domain ($\sim 3200$ dunes). These curves show that $\alpha=2$ (surface weighting) gives roughly the same weight to small and large dunes. 

Finally, as in figure~\ref{histos}b, one can compute the distribution $P_{\rm log}$ of the dune width logarithm. This is particularly interesting if one wants to test the existence of scale-free multiplicative processes, whose signature is a log-normal law. For a weighting in number, one obtains the relation: $P_{\rm log}(\ln W) d\ln W =P_0(W) dW$ so that $P_{\rm log}(\ln W) \propto W P_0(W)\propto P_1(W)$. In practice, $P_{\rm log}$ is obtained directly by bining $\ln W$ instead of $W$.

%\newpage

%___________________________________________________________________

%\end{article}
\end{document}